\documentclass[a4paper,fleqn]{cas-sc}
\usepackage[numbers,sort&compress]{natbib} 
\usepackage{tikz}
\usepackage{pgfplots}
\pgfplotsset{compat=1.17}
\usepackage{pgfplotstable}
\usetikzlibrary{plotmarks}
\usepackage{graphicx}
\graphicspath{{./}{Images/}{Latex/Images/}{Latex/}} 
\usepackage{epstopdf}

\newlength{\tempheight}
\newlength{\tempwidth}

\newcommand{\rowname}[1]
{\rotatebox{90}{\makebox[\tempheight][c]{\textbf{#1}}}}

\newcommand{\columnname}[1]
{\makebox[\tempwidth][c]{\textbf{#1}}}

\begin{document}
\let\WriteBookmarks\relax
\def\floatpagepagefraction{1}
\def\textpagefraction{.001}

\shorttitle{Attention-guided super-resolution of 4D flow MRI in carotid arteries}

\shortauthors{A.~Mokhtari and D.~Obrist}

\title [mode = title]{Attention-Guided Multi-Scale Feature Extraction Network for Super-Resolution of 4D Flow MRI in Carotid Arteries}

\author[1]{Ali Mokhtari}
\cormark[1]

\ead{ali.mokhtari@unibe.ch}

\affiliation[1]{{ARTORG Center for Biomedical Engineering Research,}, {University of Bern,},city={Bern},country={Switzerland}}

\author%
[1]
{Dominik Obrist}
\cortext[cor1]{Corresponding author}

\begin{abstract}
Four-dimensional (4D) flow magnetic resonance imaging (MRI) is a powerful non-invasive technique for visualizing and quantifying complex blood flow patterns in vivo. Despite its clinical promise, broader adoption is limited by low spatial resolution and sensitivity to noise, which restrict accurate assessment of critical hemodynamic biomarkers such as wall shear stress, pressure gradients, and turbulent kinetic energy. To overcome these challenges, we propose a deep learning-based super-resolution framework that integrates multi-scale feature extraction and attention mechanisms to enhance the quality of 4D flow MRI data.

The model was trained on a dataset of 120 patients with 240 stenosed carotid arteries. High-resolution ground truth data were generated using patient-specific computational fluid dynamics (CFD) simulations based on segmented vascular geometries and physiologically realistic boundary conditions, and the resulting velocity fields served as targets for supervised learning. The proposed architecture uses convolutional block attention modules (CBAM) to guide the network toward clinically relevant spatial features and to suppress noise in low-resolution inputs.

Quantitative results show that the attention-guided model substantially reduces the root mean square error (RMSE) compared with a baseline model without attention, and qualitative velocity contour analysis confirms improved reconstruction of intricate flow patterns. These findings highlight the capacity of the model to restore high-fidelity flow fields under noisy conditions and support the use of deep learning to extend the clinical utility of 4D flow MRI for non-invasive hemodynamic assessment.
\end{abstract}

\makeatletter
\providecommand\printorcid{}
\providecommand\printtwitter{}
\providecommand\printfacebook{}
\providecommand\printlinkedin{}
\providecommand\printemails{}
\providecommand\printurls{}
\makeatother
\maketitle

\section{Introduction}
Cardiovascular diseases are the leading cause of death globally, with ischemic stroke being one of the most devastating outcomes, often linked to pathologies of the carotid arteries~\cite{martin2025heart, roth2015estimates}. The carotid arteries, which supply oxygenated blood to the brain, are particularly prone to atherosclerotic plaque formation, leading to stenosis and increased risk of cerebrovascular events~\cite{u2009carotid}. Studying blood flow in these arteries is crucial for understanding the hemodynamic forces involved in plaque progression and embolism risk, thereby improving patient-specific diagnosis and risk stratification.

Medical imaging plays a pivotal role in diagnosing and monitoring vascular diseases. Modalities such as ultrasound, computed tomography angiography (CTA), and magnetic resonance imaging (MRI) are routinely used to assess carotid artery geometry and pathology~\cite{u2009carotid, saba2018carotid}. However, traditional imaging techniques primarily provide anatomical information and often fail to capture the complex hemodynamics that are critical for evaluating disease progression or for treatment planning.

Four-dimensional (4D) flow MRI has emerged as an advanced non-invasive modality capable of capturing time-resolved, three-directional blood flow velocities within volumetric vascular regions~\cite{dyverfeldt20154d}. Unlike conventional phase-contrast MRI or Doppler ultrasound, which are limited by angle dependence or one-directional flow encoding, 4D flow MRI provides comprehensive insights into pulsatile and multidirectional blood flow patterns. This capability is especially important in the carotid arteries, where disturbed flow, recirculation zones, and wall shear stress play a pivotal role in plaque development~\cite{markl20124d}. It also enables the assessment of intricate flow phenomena such as turbulence and vortex formation. A detailed understanding of blood flow is essential for assessing congenital heart defects, vascular disease, and conditions such as aortic aneurysms, and it supports treatment planning~\cite{dyverfeldt20154d, bissell20234d}.

\subsection{Motivation}
While 4D flow MRI offers significant advantages for examining intricate hemodynamics, it also presents several challenges in post-processing and data fidelity. These limitations include low spatial and temporal resolution, spatial averaging, partial volume effects, velocity aliasing, and phase offset errors~\cite{dyverfeldt20154d, markl20124d, stalder2008quantitative}. Such artifacts can hinder the accurate quantification of hemodynamic biomarkers and reduce the clinical utility of 4D flow data. In particular, spatial averaging, an inherent consequence of voxel-based acquisition, can lead to smoothing of fine flow features and underestimation of peak velocities, especially in regions with high spatial velocity gradients~\cite{markl20124d, casciaro20214d}. Partial volume effects, resulting from the limited voxel size relative to vessel diameter, can further confound velocity measurements near vessel boundaries~\cite{stankovic20144d}.

Moreover, low signal-to-noise ratio (SNR) in phase-contrast MRI can introduce uncertainty into the velocity data, especially in smaller vessels or in low-flow conditions~\cite{stalder2008quantitative}. Although velocity aliasing remains a concern, particularly when the velocity encoding parameter (VENC) is not optimally chosen, its impact must be considered alongside these broader acquisition and reconstruction challenges. Collectively, these limitations degrade the fidelity of derived parameters, such as wall shear stress and pressure gradients, which rely on accurate spatial derivatives and high-resolution flow-field data~\cite{cibis2016effect, donati2015non}.

To address these problems, super-resolution approaches have been investigated to augment the spatial resolution of 4D flow MRI~\cite{ferdian20204dflownet, shit2022srflow}, thereby enhancing the accuracy of velocity measurements and minimizing noise. The incorporation of 4D flow MRI into clinical practice requires ongoing research and advances in more powerful super-resolution algorithms.

\subsection{Related work}
Deep learning-based super-resolution has become an attractive way to enhance the spatial detail of 4D flow MRI and to reduce velocity noise. Ferdian et al.~\cite{ferdian20204dflownet} introduced 4DFlowNet, a residual super-resolution network trained on synthetic 4D flow MRI derived from CFD simulations. The network processes anatomical and velocity information along separate paths and reconstructs denoised high-resolution velocity fields that clearly outperform conventional interpolation, particularly in low-velocity regions. Long et al.~\cite{long2023super} adapted 4DFlowNet to the quantification of aortic regurgitation by enlarging the input and output patch dimensions so that smaller cardiovascular structures around the aortic valve are better detected, and by exploring dense blocks and cross-stage partial blocks in place of the original residual blocks. Ferdian et al.~\cite{ferdian2023cerebrovascular} later extended the approach to the cerebrovasculature, combining deep learning-based resolution enhancement with physics-informed image processing to quantify intracranial velocity, flow, and relative pressure. Rutkowski et al.~\cite{rutkowski2021enhancement} followed a closely related strategy and trained a convolutional neural network on CFD simulations of cerebrovascular geometries to enhance in vivo 4D flow MRI velocity fields. SRflow~\cite{shit2022srflow} combined super-resolution and denoising in a single network and was substantially faster and more accurate than cubic B-spline interpolation.

Physics-informed approaches embed the governing equations of fluid flow into the learning problem. Kissas et al.~\cite{kissas2020machine} used physics-informed neural networks (PINNs) to predict arterial blood pressure from 4D flow MRI data. Fathi et al.~\cite{fathi2020super} proposed a physics-informed deep neural network for the super-resolution and denoising of 4D flow MRI, in which the fluid flow physics acts as a regularization term in the loss function and the network is trained to predict flow velocities, pressure, and MRI magnitude from the complex-valued Cartesian images. Kalajahi et al.~\cite{kalajahi2025input} proposed an Input Parameterized Physics-Informed Neural Network (IP-PINN) to improve the spatio-temporal resolution of 4D flow MRI while alleviating noise, velocity aliasing, and phase errors; a convolutional neural network transforms the region of interest into latent vectors from which velocity, pressure, and spin density are predicted by a multi-layer perceptron. Saitta et al.~\cite{saitta2024implicit} proposed an unsupervised alternative based on implicit neural representations that jointly denoises and super-resolves individual 4D flow MRI acquisitions without requiring paired high-resolution training data.

Temporal super-resolution is also attracting interest, as low temporal resolution may prevent accurate capture of transient flow fluctuations. Callmer et al.~\cite{callmer2025deep} adapted spatial super-resolution networks to the temporal dimension and showed that the resulting network denoised and temporally upsampled velocity data better than conventional interpolation techniques.

Most of these networks rely on CFD simulations of idealized geometries or of a small number of patient-specific geometries to provide high-resolution training targets. In the present work, we build on this paradigm with a large cohort of 240 patient-specific carotid artery simulations and combine multi-scale feature extraction with channel and spatial attention to improve the robustness of the reconstruction to measurement noise.

\subsection{Contributions}
In this study, we developed a structured pipeline to enhance the spatial resolution of 4D flow MRI for carotid artery analysis. We first assembled a low-resolution 4D flow MRI dataset from 120 patients with stenosed carotid arteries, which served as the foundation for further refinement. Patient-specific CFD simulations were then used to generate a high-resolution dataset capturing hemodynamic patterns in detail. Given the high resolution of the CFD output, the simulated velocity fields were interpolated onto a grid with twice the resolution of the original MRI data. Finally, we applied our attention-based super-resolution model to overcome the base model's limitations in predicting complex flow fields, thereby bridging the gap between the initial imaging constraints and the level of detail required for advanced hemodynamic analysis. Our contributions can be summarized as follows:

\begin{enumerate}
\item
Curating a 4D flow MRI dataset of 120 patients with 240 stenosed carotid arteries, forming a reliable baseline for further enhancements.
\item
Generating a high-resolution dataset using patient-specific CFD simulations, providing detailed hemodynamic information beyond standard imaging capabilities.
\item
Developing an attention-based super-resolution model that overcomes the inability of the base model to accurately predict complex flow fields, leading to improved vascular imaging.
\end{enumerate}

\section{Methods}
\subsection{Study population}
The study cohort was drawn from a previously published dataset~\cite{strecker2020carotid}. Patients were included if they had an internal carotid artery plaque of at least 1.5~mm thickness with a degree of stenosis below 50\% (NASCET criteria). In total, the cohort comprised 120 patients (86 males and 34 females) with 240 stenosed carotid arteries. The mean degree of stenosis was $34 \pm 17\%$.

\subsection{MRI measurements}
4D flow MRI data were acquired using a 3T MRI scanner (Prisma, Siemens Healthineers, Erlangen, Germany) equipped with an 8-channel surface coil (NORAS MRI Products GmbH, Hoechberg, Germany) at the Department of Neurology and Neurophysiology, Freiburg University Hospital. 4D flow MRI acquisition was performed using a prospectively ECG-triggered k-t-accelerated 3D phase-contrast sequence, achieving an isotropic spatial resolution of 0.8 mm and a temporal resolution of 52.8 ms. The imaging parameters are detailed in Table~\ref{tab:4Dflow_params}.
\begin{table}[h]
    \centering
    \caption{Imaging parameters of the 4D flow MRI acquisition.}
    \label{tab:4Dflow_params}
    \begin{tabular}{l c}
        \hline
        Parameter & Value \\
        \hline
        Spatial resolution & 0.8 mm isotropic \\
        Temporal resolution & 52.8 ms \\
        TR/TE & 52.8/3.9 ms \\
        Flip angle & 12$^\circ$ \\
        GRAPPA acceleration factor & 5 \\
        Field of view (FOV) & 140 × 140 mm$^2$ \\
        Slice thickness & 0.8 mm \\
        Bandwidth (BW) & 460 Hz/Px \\
        VENC (in-plane) & 0.6 m/s \\
        VENC (through-plane) & 1.0 m/s \\
        \hline
    \end{tabular}
\end{table}

The input data comprised patient-specific 4D flow MRI, including both magnitude and phase images, which were fully anonymized and processed in accordance with the Declaration of Helsinki. Preprocessing was performed using the custom-made extension CaroTo of the MEVISFlow research software (Fraunhofer MEVIS, Bremen, Germany)~\cite{Wehrum2014} and included noise filtering, eddy-current correction, and velocity-aliasing mitigation. For segmentation, an nnU-Net~\cite{isensee2021nnu} was trained using semi-manually generated labels annotated by an experienced neurologist within the CaroTo framework. The segmentation model achieved a median Dice similarity coefficient of 0.88 on the test dataset. In the subsequent step, the phase images were masked and transformed into velocity vector fields.

\subsection{Patient-specific CFD simulation workflow}

Patient-specific vascular anatomies were reconstructed from the segmented MRI data and exported as STL (stereolithography) files. Prior to the CFD analysis, a Taubin smoothing filter was applied to reduce surface irregularities while preserving critical geometric details~\cite{taubin1995smoothing}.

Centerlines were subsequently extracted using the Vascular Modeling Toolkit (VMTK) (\href{http://www.vmtk.org}{www.vmtk.org}). The extraction algorithm identified the minimum-cost path, where cost was defined as the inverse of the largest inscribed sphere's radius along the vessel centerline. Boundary planes, oriented normal to the centerline, were then generated to facilitate downstream meshing.

High-quality meshes were created using \textit{blockMesh} and \textit{snappyHexMesh} within the OpenFOAM framework, customized to accommodate anatomical variability across patients. A maximum base cell size of 0.05~mm was set to ensure resolution of the smallest eddies~\cite{stroud2002numerical}. Mesh refinement by a factor of two was applied locally within the carotid bulb and stenotic segments to accurately resolve complex flow structures. Five boundary layers with a growth factor of 1.2 and an initial layer thickness of 0.01 mm were incorporated to effectively capture near-wall gradients.

Physiological boundary conditions were imposed based on phase-contrast MRI (PC-MRI) velocity fields~\cite{bozzi2017uncertainty}. Prior to mapping MRI data to the CFD mesh, rigid registration was performed using an Iterative Closest Point (ICP) algorithm to align the imaging and CFD geometries~\cite{zhang2021icp}. MRI-derived velocities were spatially interpolated onto the finer CFD mesh and temporally interpolated across cardiac phases to maintain continuous inlet conditions throughout the cardiac cycle.

Due to vascular compliance, side branches, and inherent uncertainties in PC-MRI measurements, discrepancies between instantaneous inflow and outflow rates may arise. To address this, flow rates were computed at four equidistant cross-sections along each branch, averaged, and used to correct outlet conditions. Flow waveforms of the CCA, ICA, and ECA were temporally synchronized. The ICA and ECA flow rates were proportionally scaled relative to the CCA waveform while maintaining the physiological ICA:ECA ratio~\cite{hoi2010effect}. The corrected flow waveform was applied at the ICA outlet, whereas a stress-free boundary condition was imposed at the ECA outlet~\cite{morbiducci2010outflow}. Vessel walls were modeled as rigid with no-slip velocity conditions; the initial velocity field was set to zero.

Simulations were conducted under a quasi-direct numerical simulation (q-DNS) approach, resolving the Navier--Stokes equations without turbulence models but employing second-order spatial and temporal discretization for computational efficiency~\cite{komen2014quasi}. All simulations were performed using OpenFOAM~\cite{weller1998tensorial}, with pressure-velocity coupling handled via the PIMPLE algorithm~\cite{holzmann2019mathematics}. Blood was assumed Newtonian, with a viscosity of $\mu = 0.004$~Pa$\cdot$s and a density of $\rho = 1060$~kg/m$^3$. An adaptive time-stepping strategy ensured a Courant--Friedrichs--Lewy (CFL) number below 0.6, balancing temporal resolution with computational stability.

For each patient case, simulations were run for three complete cardiac cycles. To minimize the impact of initial conditions on the final results, data from the initial two cycles were excluded from the subsequent analysis~\cite{lee2008direct}. All q-DNS simulations were executed on the Eiger system, part of the ALPS infrastructure at the Swiss National Supercomputing Center (CSCS), utilizing 256 processing cores per simulation.

\section{Proposed Architecture}

\subsection{Base model architecture}
\begin{figure}[ht]
\centering
\includegraphics[scale=0.5, trim=0cm 3cm 0cm 2cm, clip=true]{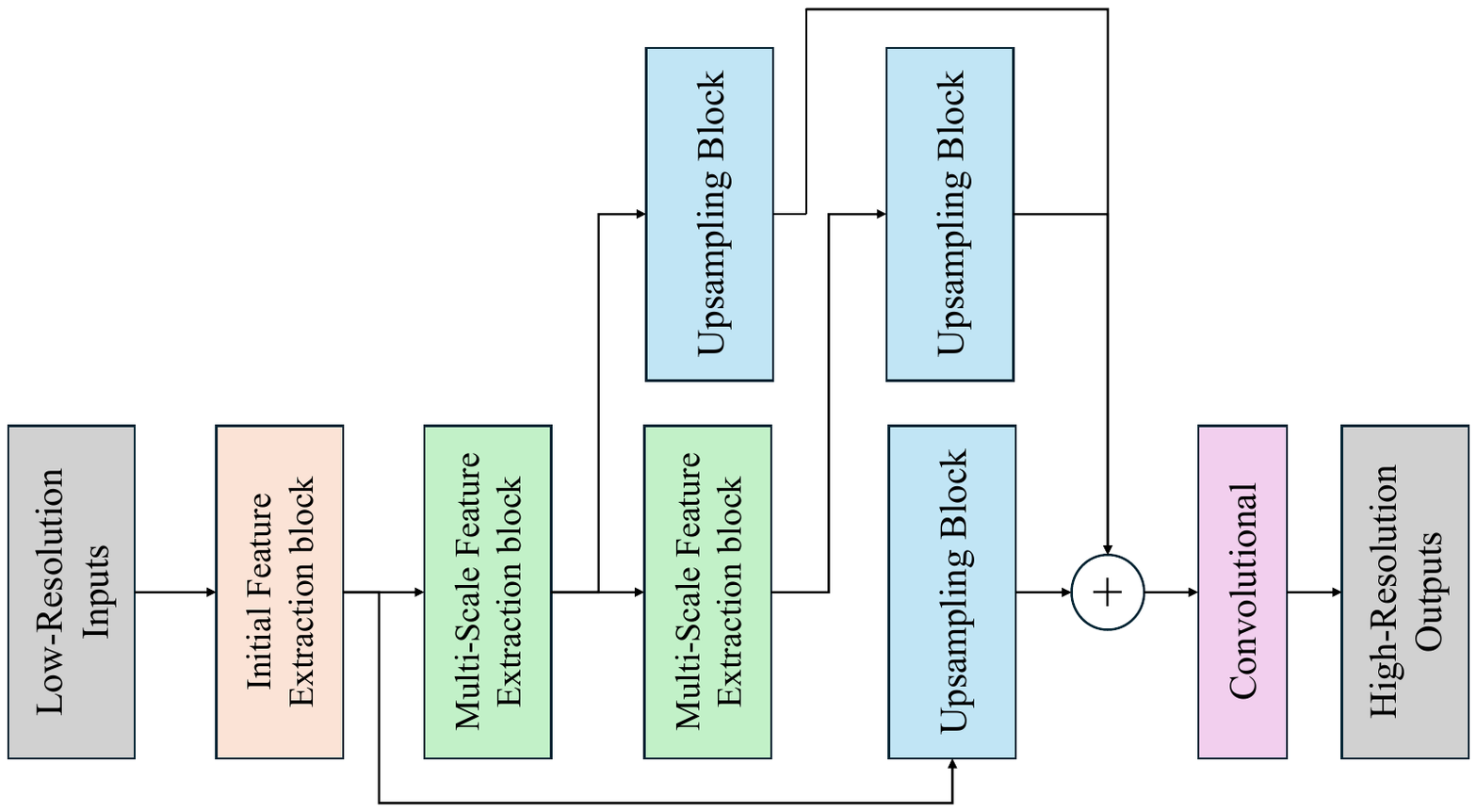}
\caption{Overview of the base super-resolution architecture.}
\label{model_base}
\end{figure}

We present a deep learning architecture for 3D super-resolution that enhances the spatial resolution of volumetric velocity data. The architecture takes low-resolution inputs and progressively refines the extracted features to produce high-resolution outputs. As illustrated in Figure~\ref{model_base}, it consists of an initial feature extraction (IFE) block, two multi-scale feature extraction (MSFE) blocks arranged in sequence, three upsampling blocks, and a final convolutional layer. The IFE block identifies essential patterns and structures in the incoming data. The two MSFE blocks then analyze the features at different scales, thereby preserving both fine details and broader contextual information. The output of each feature extraction block is routed along two pathways: one continues to the next block for further refinement, while the other is passed through an upsampling block that doubles the spatial dimensions. The upsampled feature maps are summed, and a final convolutional layer consolidates the fused information into a high-resolution output with twice the spatial resolution of the input in each direction. The two main components, the IFE block and the MSFE block, are explained below.

\subsubsection{Initial feature extraction block}
\begin{figure}[ht]
\centering
\includegraphics[scale=0.25, trim=0cm 3cm 0cm 2cm, clip=true]{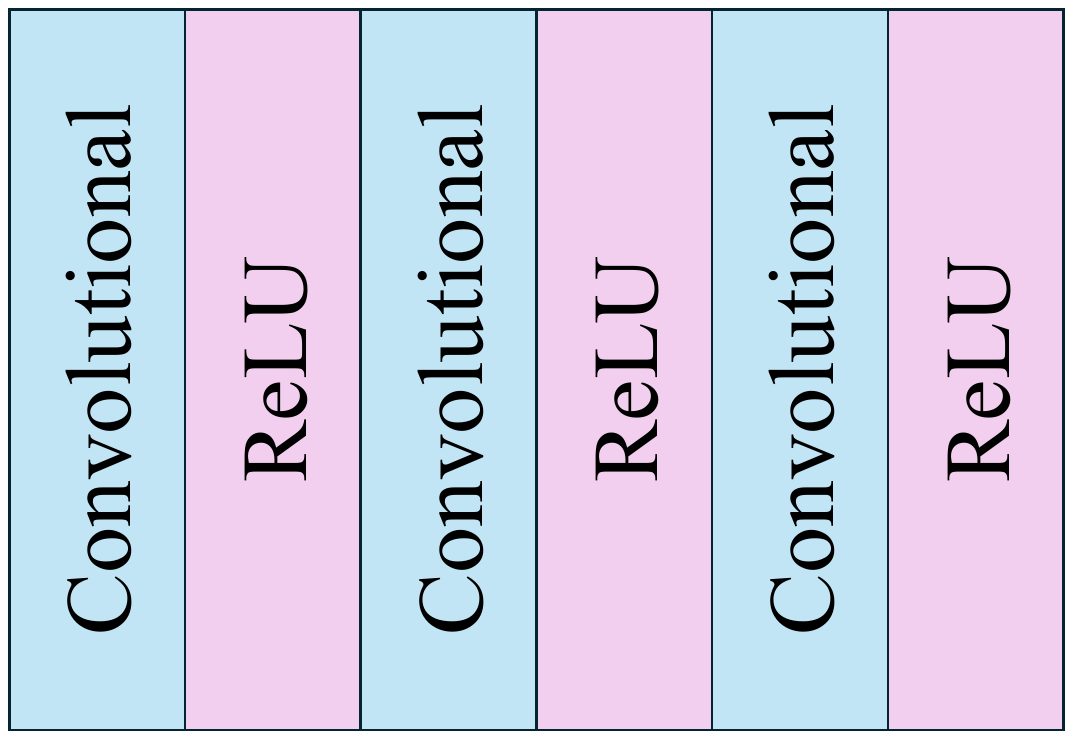}
\caption{The Initial Feature Extraction block.}
\label{ife}
\end{figure}

Figure~\ref{ife} shows the Initial Feature Extraction (IFE) block, which forms the foundation of the architecture. It comprises three 3D convolutional layers with ReLU activations that gradually extract features from the input volumetric data, increasing feature dimensionality while preserving essential spatial information.

Each upsampling block in the architecture includes a transposed 3D convolution layer that doubles the spatial dimensions.

\subsubsection{Multi-scale feature extraction block}
\begin{figure}[ht]
\centering
\includegraphics[scale=0.5, trim=0cm 6cm 0cm 5cm, clip=true]{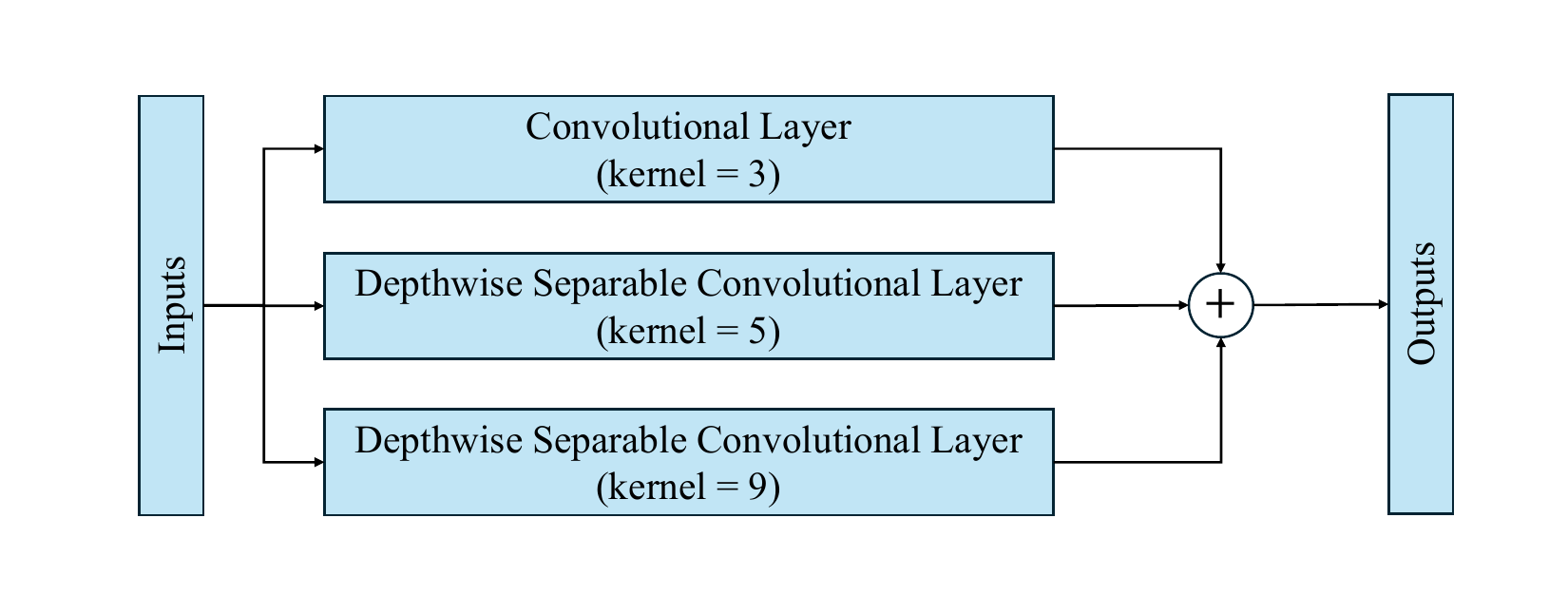}
\caption{The Multi-Scale Feature Extraction block.}
\label{msfe}
\end{figure}

The Multi-Scale Feature Extraction (MSFE) block improves feature refinement using a multi-scale convolutional operation, illustrated in Figure~\ref{msfe}. The input features are processed in parallel by a conventional 3D convolutional layer with a kernel size of 3, and by two depthwise separable 3D convolutional layers with kernel sizes of 5 and 9; the three outputs are then summed. The use of parallel convolutions with different kernel sizes enables the model to identify features at different scales, enhancing its ability to reproduce complex details and textures in the super-resolved output. Each MSFE block produces an output that flows through two distinct pathways: one is further processed by the next MSFE block to obtain refined features, while the other is upsampled by an upsampling block and used for the final combination at the end of the architecture.

Within the MSFE block, using a 3D convolutional layer with a larger kernel size significantly increases computational cost. We therefore use 3D depthwise separable convolutional layers instead of conventional 3D convolutional layers to support enlarged kernel sizes. Depthwise convolution is a distinct type of convolution that operates on each input channel independently, employing a specific filter for each channel. That is, with $C$ input channels, $C$ distinct filters are employed, each corresponding to an individual input channel. Following the depthwise convolution, a pointwise ($1 \times 1 \times 1$) convolution is applied to combine the outputs of the depthwise layer. Depthwise separable convolution offers a lower computational cost by substantially reducing the number of parameters and operations compared with conventional convolution. This reduction allows faster training and inference, making it appropriate for real-time applications. Furthermore, despite the reduced computational complexity, depthwise separable convolution can maintain or enhance model performance by efficiently capturing spatial information while mitigating the risk of overfitting. Additionally, it can be easily integrated into existing models, leading to more efficient designs without sacrificing model quality.

\subsection{Attention-based model}
Convolutional neural networks (CNNs) have significant potential for improving image quality in 4D flow MRI. Nonetheless, CNNs may insufficiently preserve essential information about blood flow, especially in regions where flow varies rapidly, despite their ability to improve image accuracy. Noise in low-resolution datasets can distort the spatial and temporal relationships between regions, resulting in inaccurate reconstructions. Furthermore, noise may obscure critical information in low-resolution images, limiting traditional CNNs' ability to discern meaningful patterns. We therefore add attention mechanisms to the super-resolution process to sharpen the reconstructed details and address these issues by more effectively selecting important features. By employing attention, the model can concentrate on the most salient features of the input, thereby efficiently diminishing noise and improving the signal-to-noise ratio. The attention mechanism helps the model weigh the importance of different parts of the input for the current task. This capability enhances the feature representation and ultimately produces more accurate super-resolution outputs.

\begin{figure}[ht]
\centering
\includegraphics[scale=0.5, trim=0cm 3cm 0cm 2cm, clip=true]{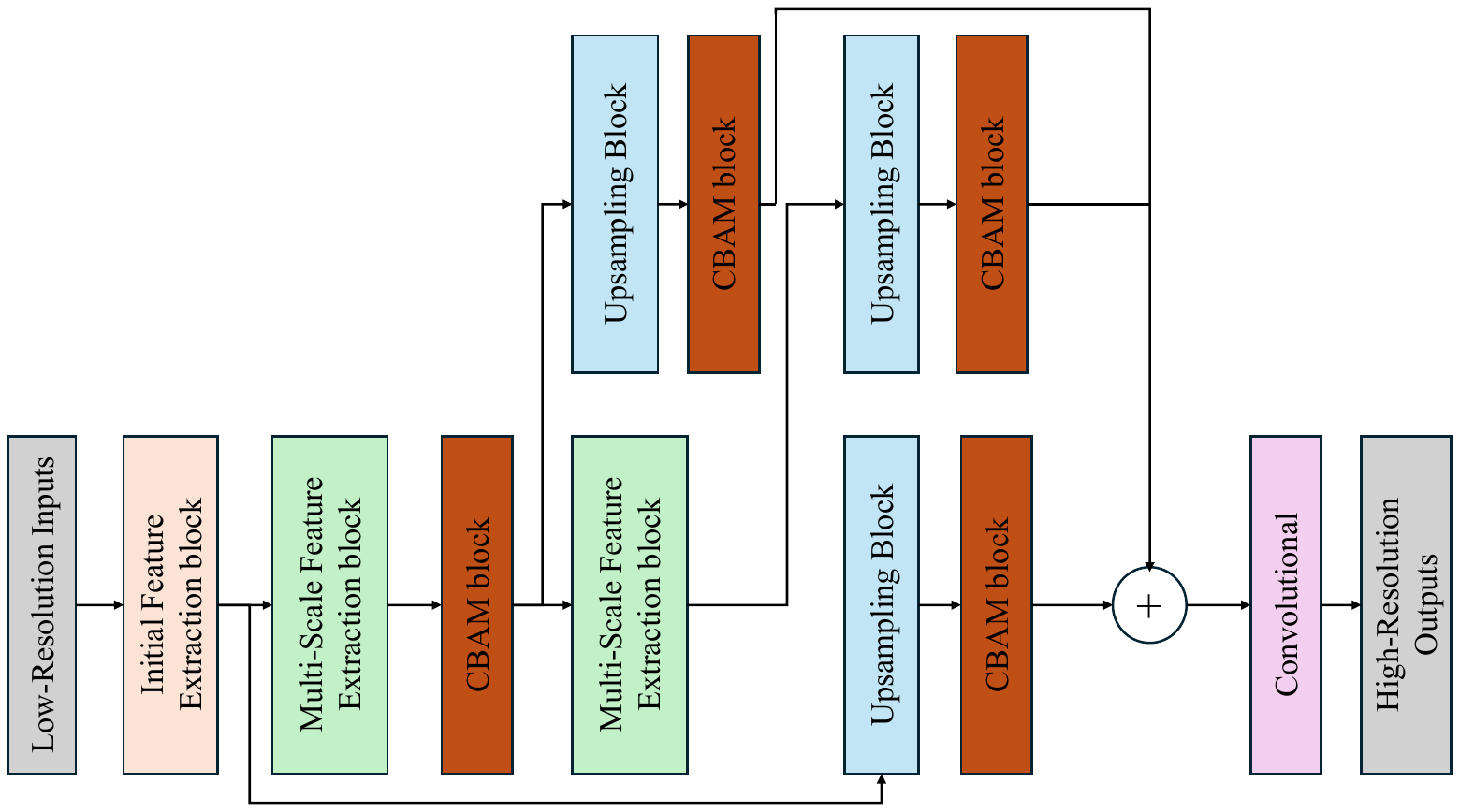}
\caption{Architecture of the attention-based model with CBAM blocks.}
\label{model_att}
\end{figure}

To achieve this, we combined channel and spatial attention, as shown in Figure~\ref{model_att}. Convolutional block attention modules (CBAM, described below) are inserted after the first MSFE block and after each of the three upsampling blocks. The attention modules suppress redundant features, and channel attention in particular is important for eliminating noisy features.

\subsubsection{Convolutional block attention module}
\begin{figure}[ht]
\centering
\includegraphics[scale=0.5, trim=1cm 6cm 1cm 4cm, clip=true]{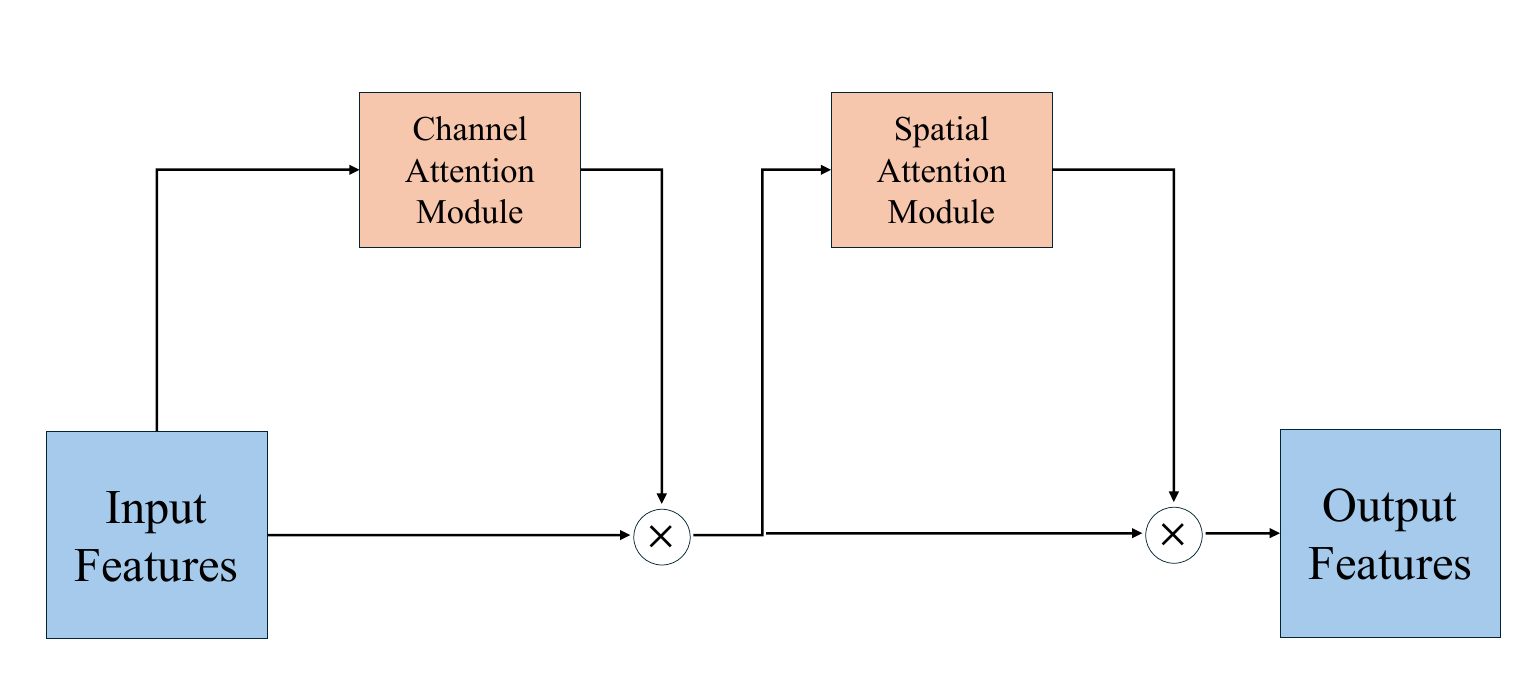}
\caption{Schematic representation of the convolutional block attention module (CBAM).}
\label{cbam1}
\end{figure}

The Convolutional Block Attention Module (CBAM) is an effective attention mechanism developed by Woo et al.~\cite{woo2018cbam} that improves the capabilities of convolutional layers, as illustrated in Figure~\ref{cbam1}. CBAM operates through two sequential steps: channel attention and spatial attention.

\begin{figure}[ht]
\centering
\includegraphics[scale=0.5, trim=1cm 5cm 1cm 4cm, clip=true]{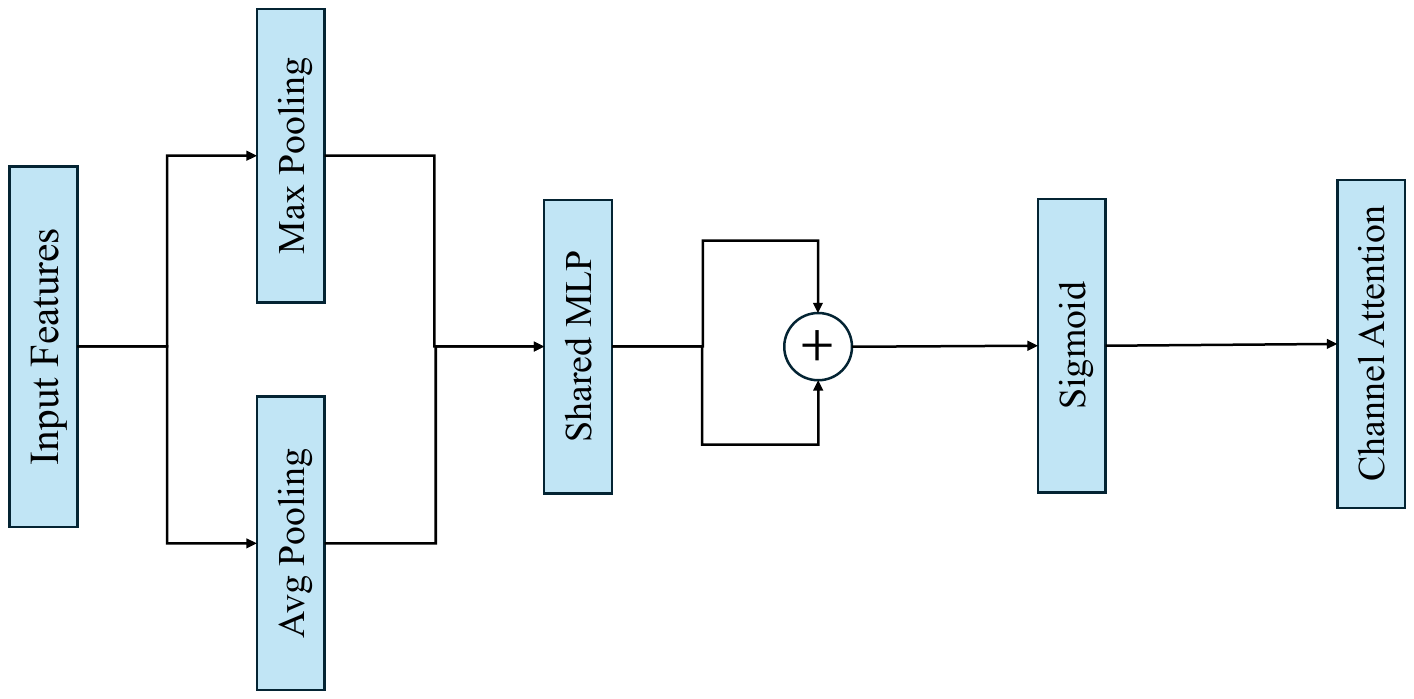}
\caption{Channel attention module.}
\label{cam}
\end{figure}

As illustrated in Figure~\ref{cam}, during the channel attention phase, the module produces a channel-specific attention map by integrating feature maps across spatial dimensions, allowing the network to focus on the most relevant channels. This is achieved by global average pooling and global max pooling, which capture different aspects of the feature distribution. The generated attention map is then applied to the original feature maps, enhancing significant channels while attenuating less critical ones.

\begin{figure}[ht]
\centering
\includegraphics[scale=0.5, trim=1cm 5cm 1cm 4cm, clip=true]{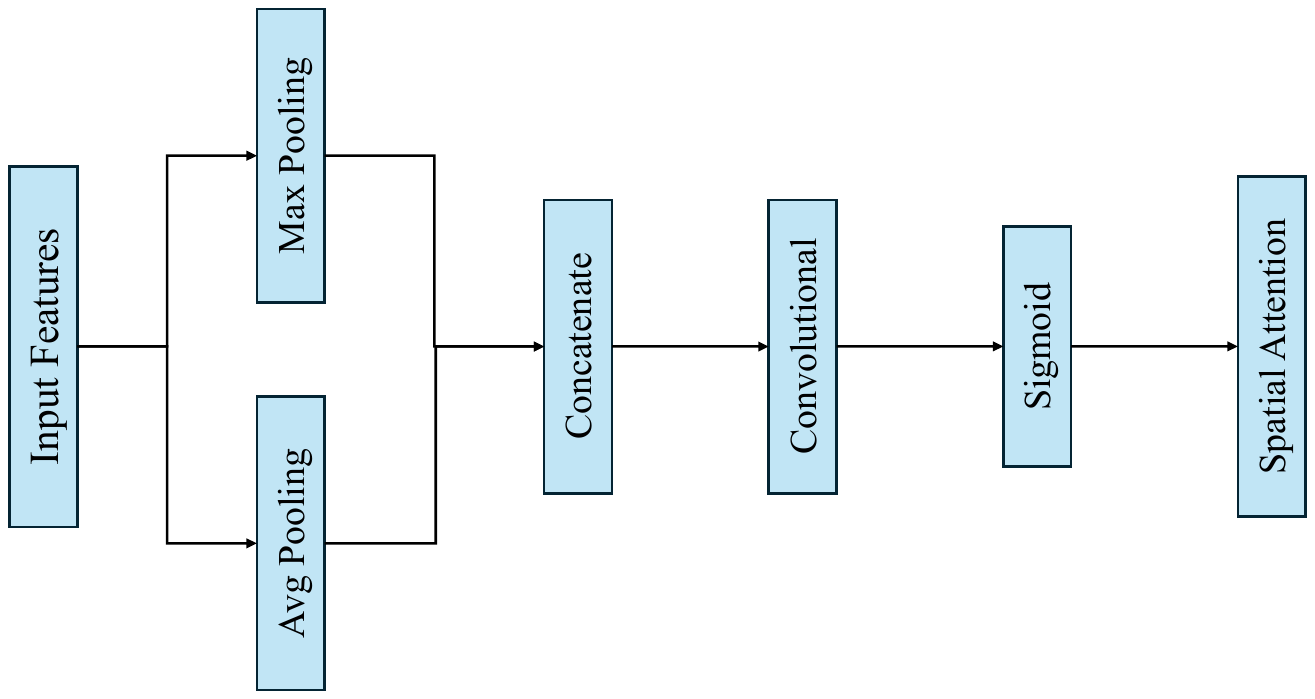}
\caption{Spatial attention module.}
\label{sam}
\end{figure}

Following channel attention, a spatial attention mechanism is employed to further enhance the feature maps by focusing on specific spatial regions, as shown in Figure~\ref{sam}. This is accomplished by creating a spatial attention map that emphasizes areas of interest within the feature maps, employing a comparable pooling method to gather information across channels. The integration of channel and spatial attention enables CBAM to substantially improve the feature representation, resulting in enhanced performance across multiple vision tasks, including image classification, object detection, and segmentation. Integrating CBAM into existing CNN designs has resulted in substantial accuracy improvements at negligible computational cost, making it a useful tool for enhancing the deep learning models of our super-resolution framework.

\section{Results \& Discussion}
Two models were trained: a base model and an attention-based model, both implemented in PyTorch and trained on a single NVIDIA H100 GPU. Training ran for at most 200 epochs, with early stopping after 15 epochs without improvement to avoid overfitting. The Adam optimizer, known for its effectiveness in addressing complex optimization problems, was used for both models, and mean squared error (MSE) was the loss function. Our dataset consisted of noisy 4D flow MRI data, which posed challenges due to its inherent variability. We compared the performance of the base model with that of the attention-based model to evaluate the influence of the attention mechanism on the model's ability to focus on relevant characteristics in the noisy dataset, thereby improving its ability to accurately identify the underlying patterns in the 4D flow MRI dataset.

\begin{figure}[ht]
\centering
\includegraphics[scale=0.7]{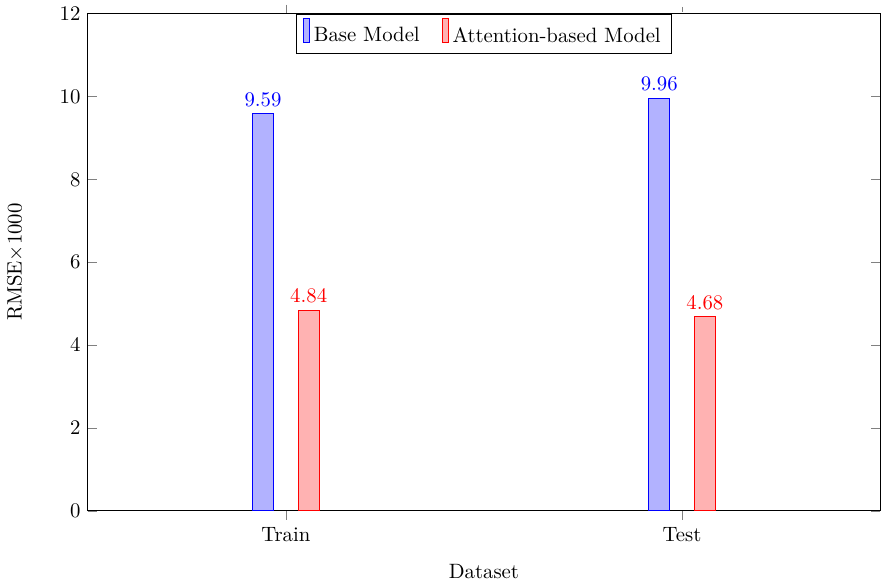}
\caption{Comparison of RMSE for the base model and the attention-based model.}
\label{rmse_bar}
\end{figure}

The root mean square error (RMSE) in Figure~\ref{rmse_bar} shows a clear difference in how well the two models perform for super-resolution of 4D flow MRI data. The base model had a higher RMSE, indicating it was less accurate at reconstructing the velocity fields, likely because it struggled with the high noise levels in the input data. In contrast, the attention-based model achieved nearly half the RMSE of the base model, showing that it is much better at reducing noise and focusing on key features.

\begin{figure}[ht]
\centering
\begin{tikzpicture}

\definecolor{darkslategray83}{RGB}{83,83,83}
\definecolor{dimgray85}{RGB}{85,85,85}
\definecolor{gainsboro229}{RGB}{229,229,229}
\definecolor{indianred2049072}{RGB}{204,90,72}

\begin{axis}[
axis line style={black},
tick align=outside,
tick pos=left,
x grid style={black},
xmin=-0.5, xmax=1.5,
xtick style={color=black},
xtick={0,1},
xticklabels={Base Model, Attention-Based Model},
y grid style={black},
ylabel=\textcolor{black}{RMSE $\times$ 100},
ymin=-0.5, ymax=4,
ytick style={color=black}
]
\path [draw=darkslategray83, fill=indianred2049072]
(axis cs:-0.4,0.67247655)
--(axis cs:0.4,0.67247655)
--(axis cs:0.4,1.0344616)
--(axis cs:-0.4,1.0344616)
--(axis cs:-0.4,0.67247655)
--cycle;
\addplot [darkslategray83]
table {%
0 0.67247655
0 0.1466149
};
\addplot [darkslategray83]
table {%
0 1.0344616
0 1.5769666
};
\addplot [darkslategray83]
table {%
-0.2 0.1466149
0.2 0.1466149
};
\addplot [darkslategray83]
table {%
-0.2 1.5769666
0.2 1.5769666
};
\addplot [black, mark=o, mark size=3, mark options={solid,fill opacity=0,draw=darkslategray83}, only marks]
table {%
0 0
0 0.0725824
0 0.0725824
0 0.0725824
0 0.0725824
0 0.0725824
0 0.0725824
0 0.0725824
0 0.0725824
0 0.0725824
0 0.0725824
0 0.0725824
0 0.0725824
0 0.0725824
0 0.0725824
0 0.0725824
0 0.0725824
0 0.0725824
0 0.0725824
0 0.0725824
0 0.0725824
0 0.0725824
0 0.0725824
0 0.0725824
0 0.0725824
0 0.0725824
0 0.0725824
0 0.0725824
0 0.0725824
0 0.0725824
0 0.0725824
0 0.0725824
0 0.0725824
0 0.0725824
0 0.0725824
0 0.0725824
0 0.0725824
0 0.0725824
0 0.0725824
0 0.0725824
0 0.0725824
0 0.0725824
0 0.0725824
0 0.0725824
0 0.0725824
0 0.0725824
0 0.0725824
0 0.0725824
0 0.0725824
0 2.1189616
0 2.0443919
0 2.079662
0 1.944108
0 1.6613915
0 2.1513914
0 2.3244379
0 2.2947513
0 2.0727843
0 1.6502047
0 1.6293245
0 1.6520664
0 1.5792087
0 2.0273289
0 2.4878487
0 2.2684285
0 2.2250836
0 2.0081512
0 2.4453234
0 2.7497094
0 2.1931843
0 2.0569469
0 1.7891727
0 3.0539253
0 3.013293
0 2.5924779
0 2.7138743
0 2.4094818
0 1.6713299
0 3.4687098
0 2.9954164
0 2.6502122
0 2.6329868
0 2.2496318
0 1.7592803
0 1.9005713
0 1.6196059
0 1.891676
0 2.0689278
0 1.6913513
0 2.6788602
0 2.4875013
0 2.2616277
0 1.8948196
0 1.5899543
0 2.03349
0 1.8610156
0 1.8230053
0 1.7996458
0 1.6409647
0 1.6786424
0 2.3126971
0 2.1839315
0 2.1721685
0 2.2392806
0 2.0958752
0 1.705638
0 1.8469281
0 1.8158181
0 1.6136335
0 1.7975054
0 1.8639775
0 1.9231397
0 1.7872711
0 2.2351367
0 3.2964447
0 2.3346534
0 1.9589837
0 1.6562543
0 2.0544387
0 3.3789941
0 2.7506673
0 2.4426187
0 2.0583468
0 2.1426662
0 1.8700496
0 1.8646516
0 1.8121819
0 2.1118397
0 2.3027006
0 2.1748464
0 1.993753
0 1.8675294
0 2.3363608
0 1.9746895
0 2.0338462
0 1.9507584
0 1.6697186
0 2.455679
0 2.3089124
0 2.3841615
0 2.2760952
0 1.8967597
0 2.5582613
0 2.3923226
0 2.4155343
0 2.0788894
0 2.3448405
0 2.148311
0 2.2000744
0 1.8616366
0 1.7542611
0 1.8953413
0 1.9069487
0 1.7478108
0 1.9354274
0 1.8106987
0 1.6429479
0 2.0864838
0 2.7517611
0 1.759316
0 1.9704817
0 2.7725218
0 2.5213291
0 2.3346013
0 2.0915737
0 1.7860621
0 2.6538268
0 2.6465807
0 2.5727241
0 2.2545302
0 1.9553599
0 1.6295095
0 1.5878622
0 1.8621123
0 2.1458054
0 1.8998247
0 1.7455402
0 2.0491065
0 2.989472
0 2.7417357
0 2.5709866
0 2.170621
0 1.630359
0 1.724958
0 1.7138058
0 1.644786
0 1.7177089
0 1.6905593
0 1.615099
0 3.0032469
0 2.6085168
0 2.4889169
0 1.8514795
0 1.6630088
};
\path [draw=darkslategray83, fill=indianred2049072]
(axis cs:0.6,0.3335695)
--(axis cs:1.4,0.3335695)
--(axis cs:1.4,0.523172)
--(axis cs:0.6,0.523172)
--(axis cs:0.6,0.3335695)
--cycle;
\addplot [darkslategray83]
table {%
1 0.3335695
1 0.0753576
};
\addplot [darkslategray83]
table {%
1 0.523172
1 0.8075254
};
\addplot [darkslategray83]
table {%
0.8 0.0753576
1.2 0.0753576
};
\addplot [darkslategray83]
table {%
0.8 0.8075254
1.2 0.8075254
};
\addplot [black, mark=o, mark size=3, mark options={solid,fill opacity=0,draw=darkslategray83}, only marks]
table {%
1 0
1 0.8780573
1 1.4115447
1 1.2142856
1 1.4694312
1 1.2436569
1 1.0694399
1 1.2118314
1 1.2961317
1 1.2582189
1 1.2588957
1 0.9378634
1 0.8971242
1 0.8254726
1 0.8153557
1 1.1324272
1 0.9940891
1 0.9402398
1 0.9556844
1 1.4835393
1 1.5754091
1 1.279122
1 1.308363
1 1.1210864
1 1.3969296
1 1.3062705
1 1.1536961
1 1.1199348
1 1.339586
1 0.8843636
1 1.3411214
1 1.296508
1 1.2317156
1 1.0389767
1 0.9677451
1 0.8686423
1 0.8354447
1 1.1215588
1 1.1372273
1 0.863744
1 0.8324104
1 1.2955838
1 1.2636756
1 1.1731444
1 0.9640285
1 0.853593
1 1.004937
1 0.9529982
1 0.8933142
1 0.8682369
1 0.8666415
1 0.8333362
1 0.8695357
1 0.898141
1 0.8574662
1 0.8290607
1 0.841282
1 0.9880669
1 0.8429242
1 0.8265406
1 0.8526438
1 0.9519446
1 1.0436398
1 1.066204
1 0.908696
1 1.2004081
1 1.7503924
1 1.3147423
1 1.1261033
1 1.0187298
1 0.9436354
1 1.3362394
1 1.1166069
1 1.0265949
1 0.9080118
1 1.1412299
1 0.9161283
1 0.9144819
1 0.971591
1 0.8182605
1 1.1395091
1 1.2652623
1 1.2135946
1 1.0307469
1 1.1041481
1 1.2027391
1 1.0405274
1 1.0678447
1 1.0815196
1 0.9894825
1 0.8419407
1 1.2107887
1 1.1258123
1 1.1665027
1 1.0848525
1 0.9130009
1 1.4804236
1 1.4516436
1 1.6134937
1 1.2824358
1 0.9107746
1 0.8567074
1 1.0507114
1 0.9651082
1 1.2366469
1 0.9926691
1 0.879799
1 1.3008225
1 1.2830262
1 1.1497928
1 1.1185554
1 1.0487609
1 0.928399
1 1.0953206
1 1.9327635
1 1.3415115
1 0.822669
1 0.8906762
1 1.2682583
1 1.1141341
1 1.0679975
1 0.9249959
1 0.845089
1 0.8138986
1 1.2694011
1 1.3199738
1 1.2883647
1 1.1736905
1 1.0900951
1 0.8347068
1 1.015084
1 1.1128853
1 1.0713072
1 0.9031097
1 1.012467
1 1.2042386
1 1.0355627
1 0.9420856
1 1.2519493
1 1.7641058
1 1.7650986
1 1.6266968
1 1.3386653
1 0.9342014
1 1.0051142
1 0.8624858
1 0.9133819
1 1.2266446
1 1.1672462
1 1.1253087
1 1.0789862
1 0.8356663
1 1.1834435
1 1.1637407
1 1.1279353
1 0.9795107
1 0.8143491
1 0.9495312
1 0.9100375
1 0.8875033
1 0.9131549
1 0.8243331
1 1.7577816
1 1.3158766
1 0.920331
1 0.8406674
};
\addplot [darkslategray83]
table {%
-0.4 0.8339452
0.4 0.8339452
};
\addplot [darkslategray83]
table {%
0.6 0.416736
1.4 0.416736
};
\end{axis}

\end{tikzpicture}
\caption{Comparative box plot of RMSE for the base model and the attention-based model.}
\label{box}
\end{figure}
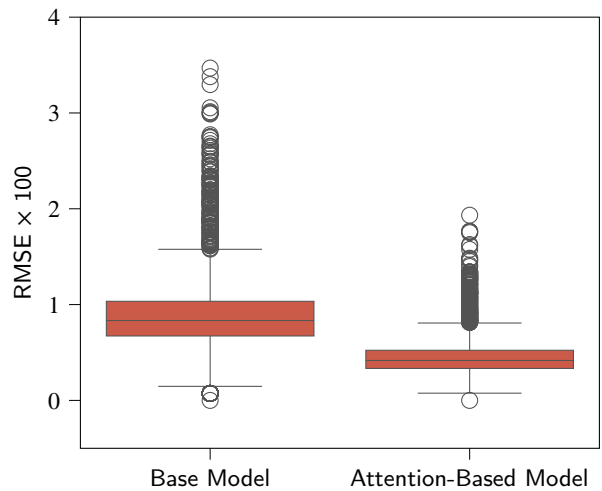

The box plots of RMSE for the base and attention-based models in Figure~\ref{box} reveal clear differences in performance across the entire dataset. The attention model exhibits a more focused distribution, suggesting greater stability, whereas the base model displays a wider interquartile range (IQR), indicating greater variance in its error distribution. Moreover, the attention model exhibits fewer extreme values, indicating more reliable predictions; conversely, the base model displays more outliers, with multiple RMSE values deviating markedly from the central range. Furthermore, the median RMSE of the attention model is lower than that of the base model, implying an overall reduction in error. These differences highlight the efficacy of the attention mechanism in enhancing predictions, reducing variance, and augmenting model reliability.

\begin{figure}[htbp]
\setlength{\tempwidth}{.2\linewidth}
\settoheight{\tempheight}{\includegraphics[width=\tempwidth]{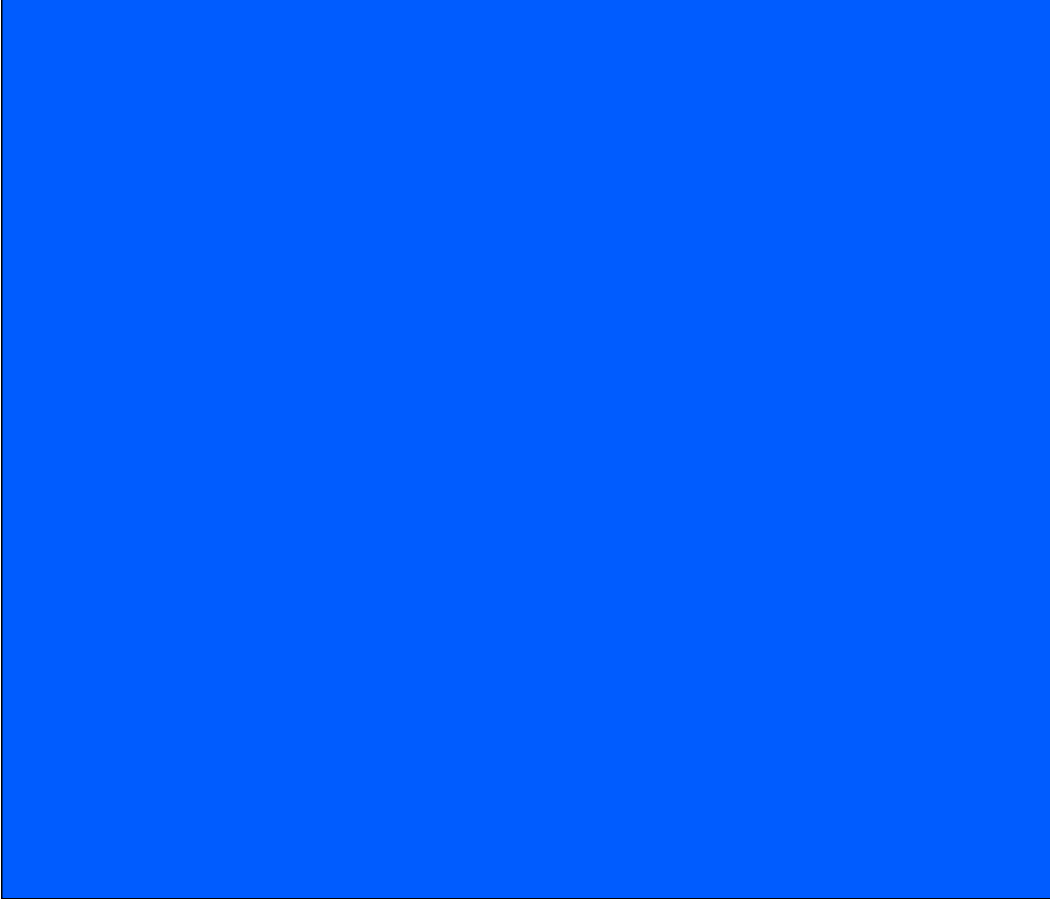}}%
\centering
\hspace{\baselineskip}
\columnname{Low resolution}\hfil
\columnname{Ground truth}\hfil
\columnname{Base model}\hfil
\columnname{Attention model}\\
\rowname{Slice 1}
{\includegraphics[width=\tempwidth]{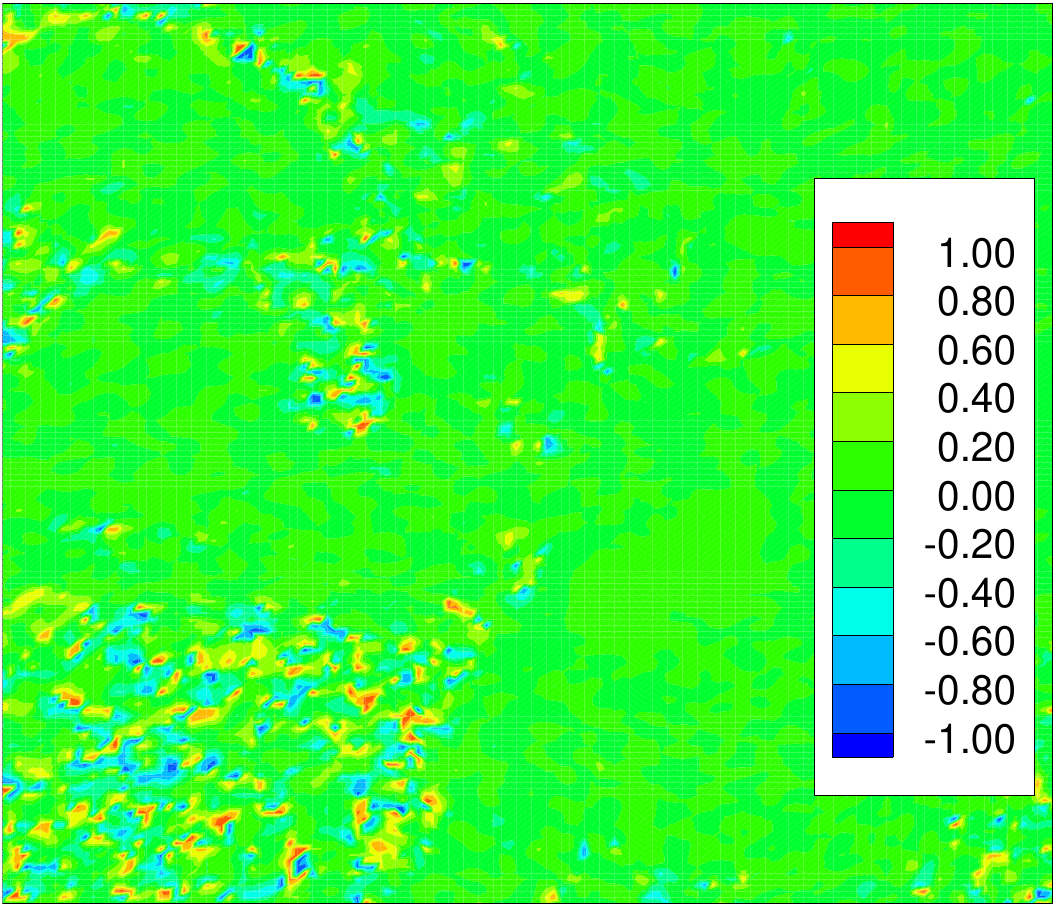}}\hfil
{\includegraphics[width=\tempwidth]{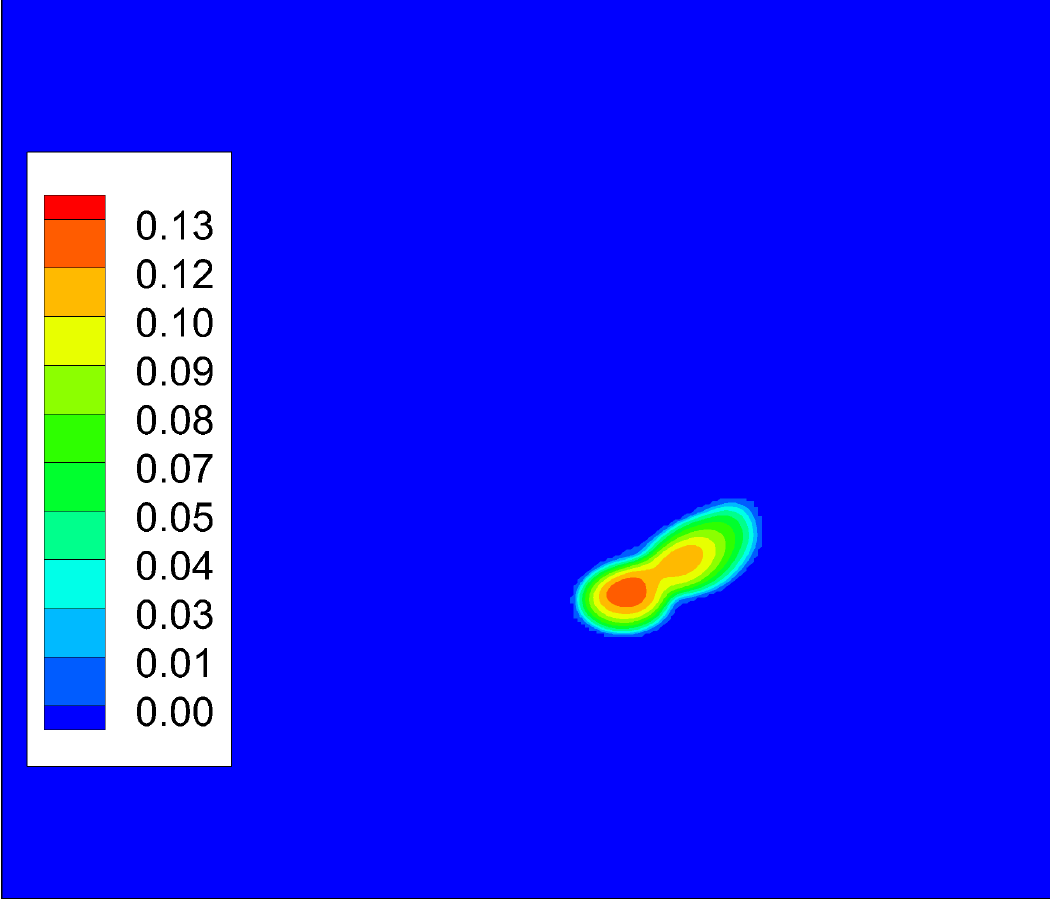}}\hfil
{\includegraphics[width=\tempwidth]{CB1}}\hfil
{\includegraphics[width=\tempwidth]{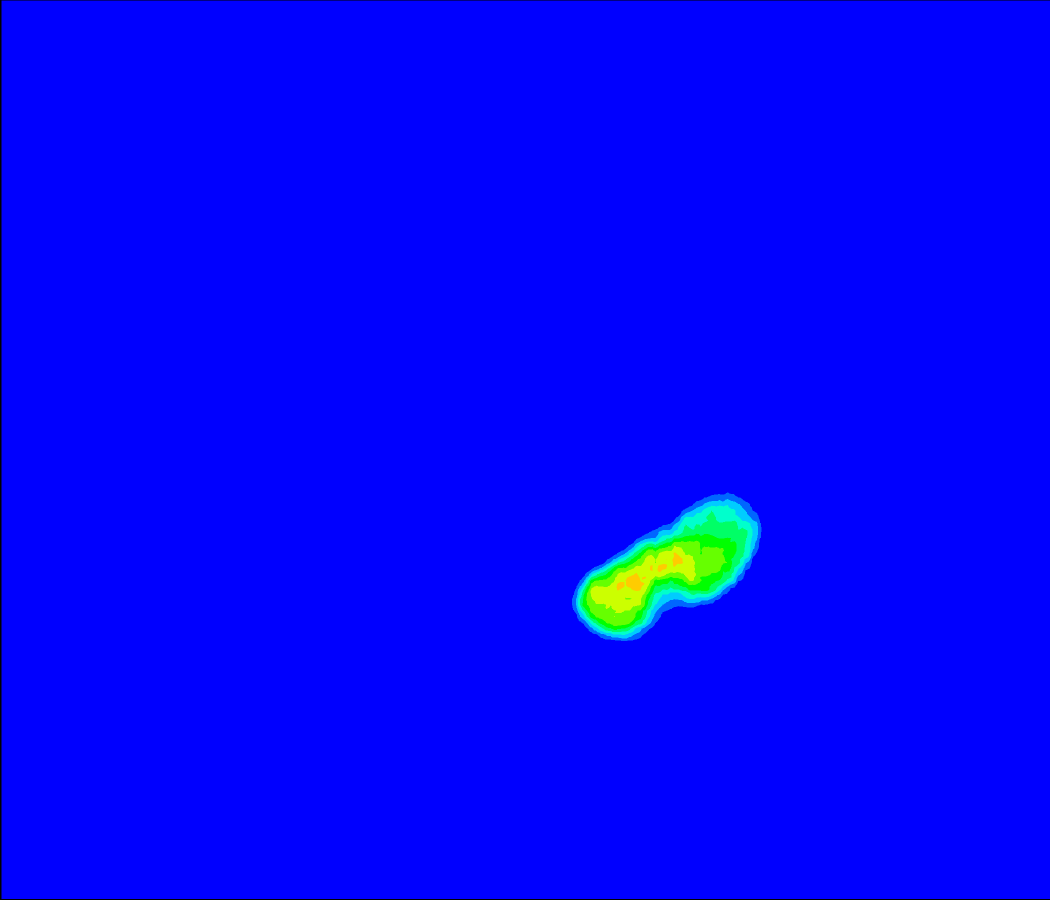}}\\
\rowname{Slice 2}
{\includegraphics[width=\tempwidth]{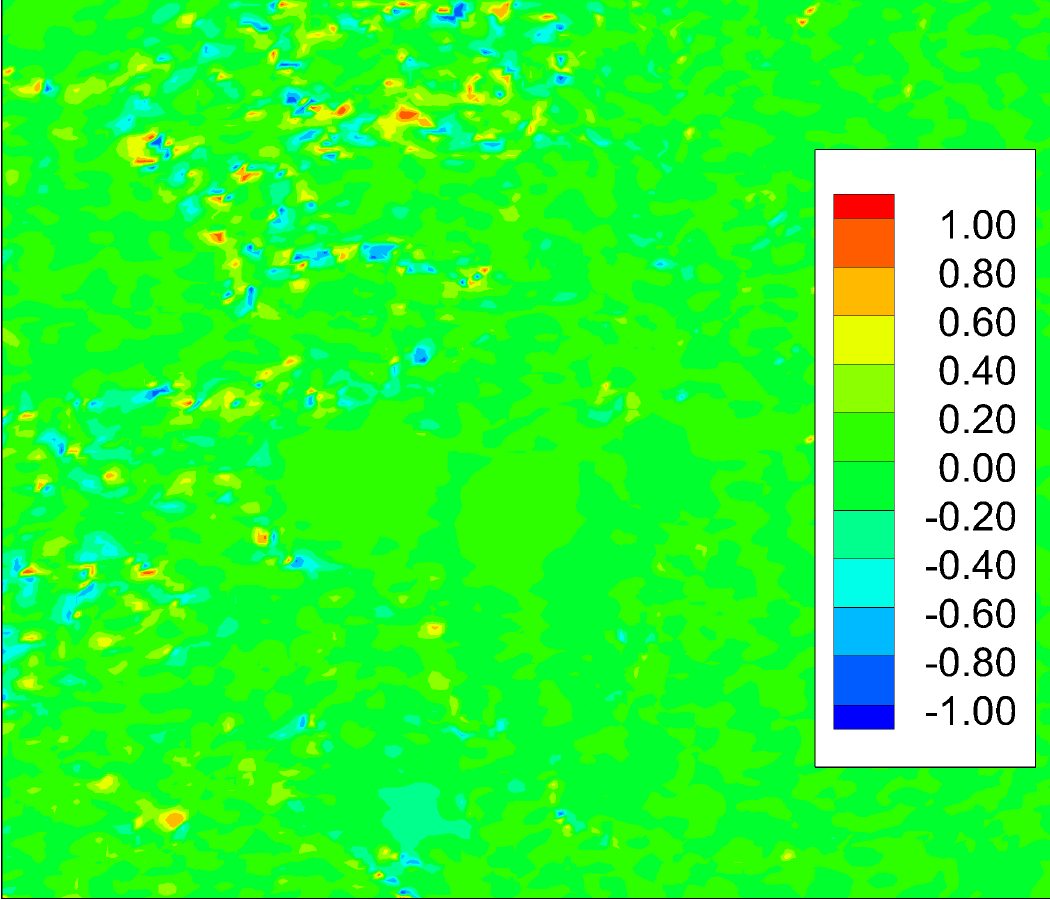}}\hfil
{\includegraphics[width=\tempwidth]{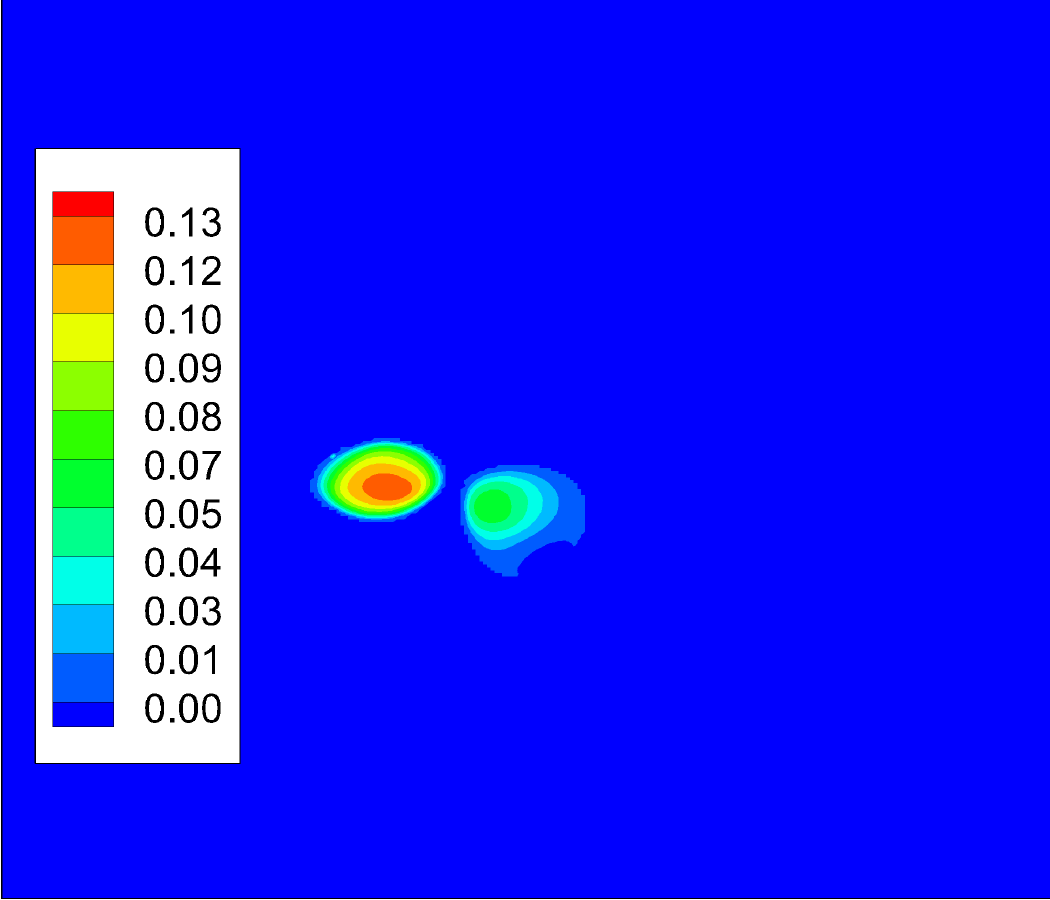}}\hfil
{\includegraphics[width=\tempwidth]{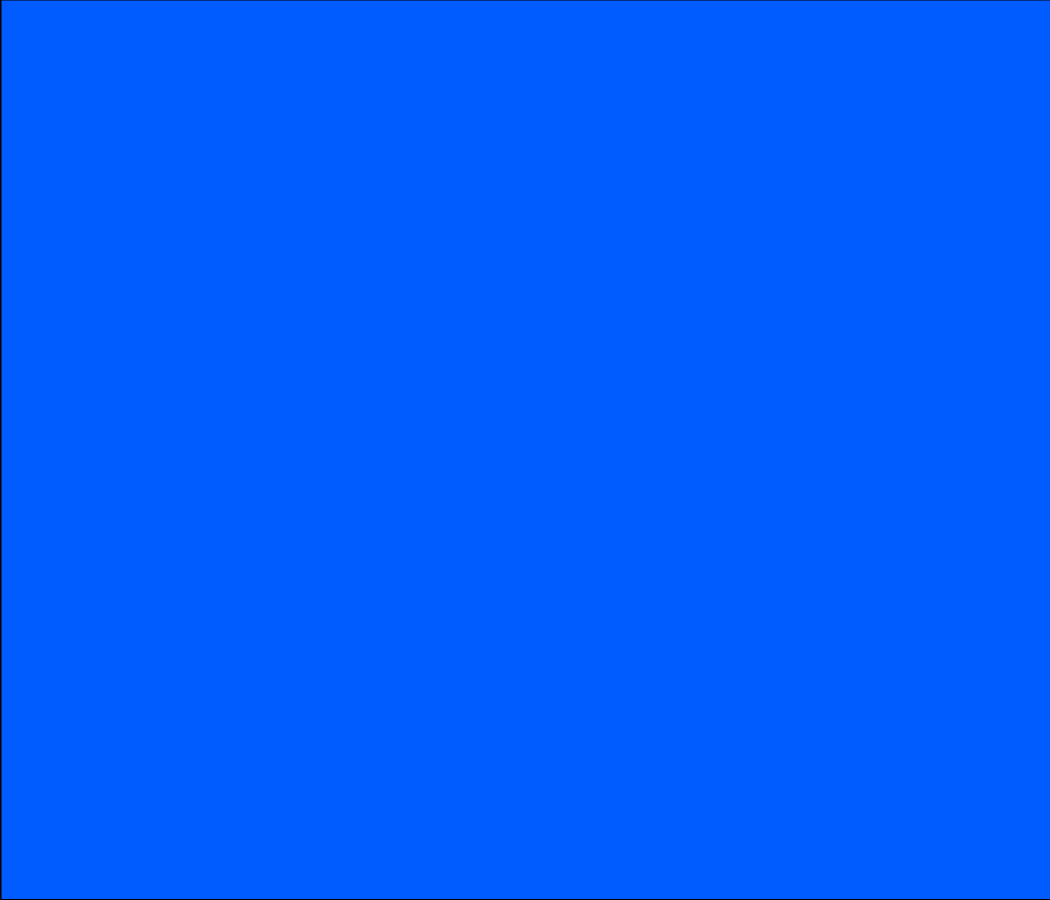}}\hfil
{\includegraphics[width=\tempwidth]{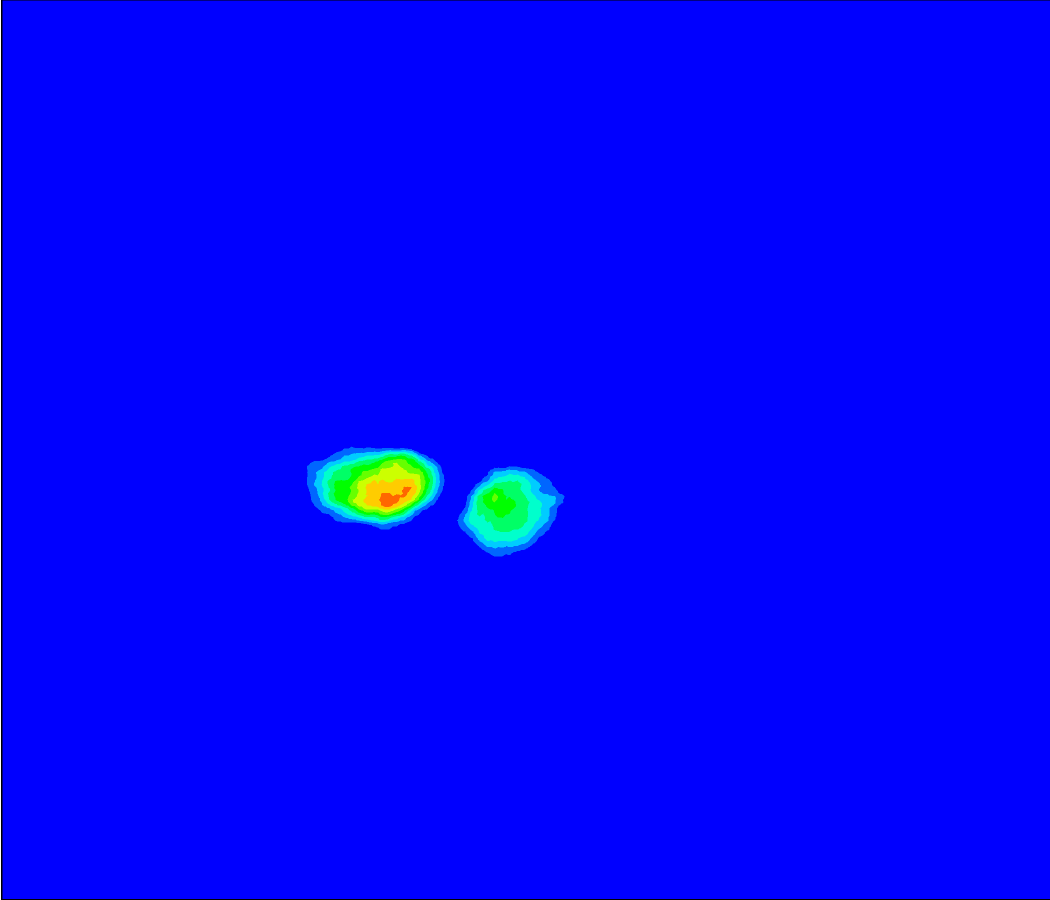}}\\
\rowname{Slice 3}
{\includegraphics[width=\tempwidth]{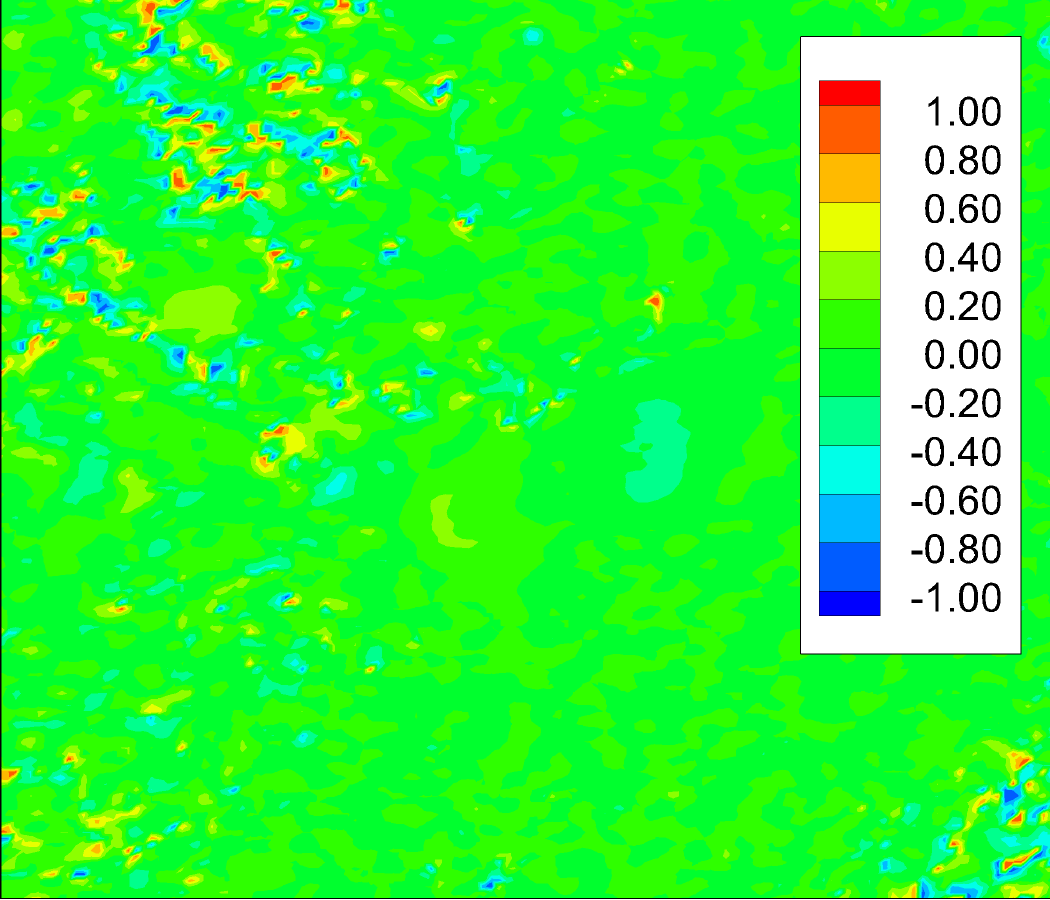}}\hfil
{\includegraphics[width=\tempwidth]{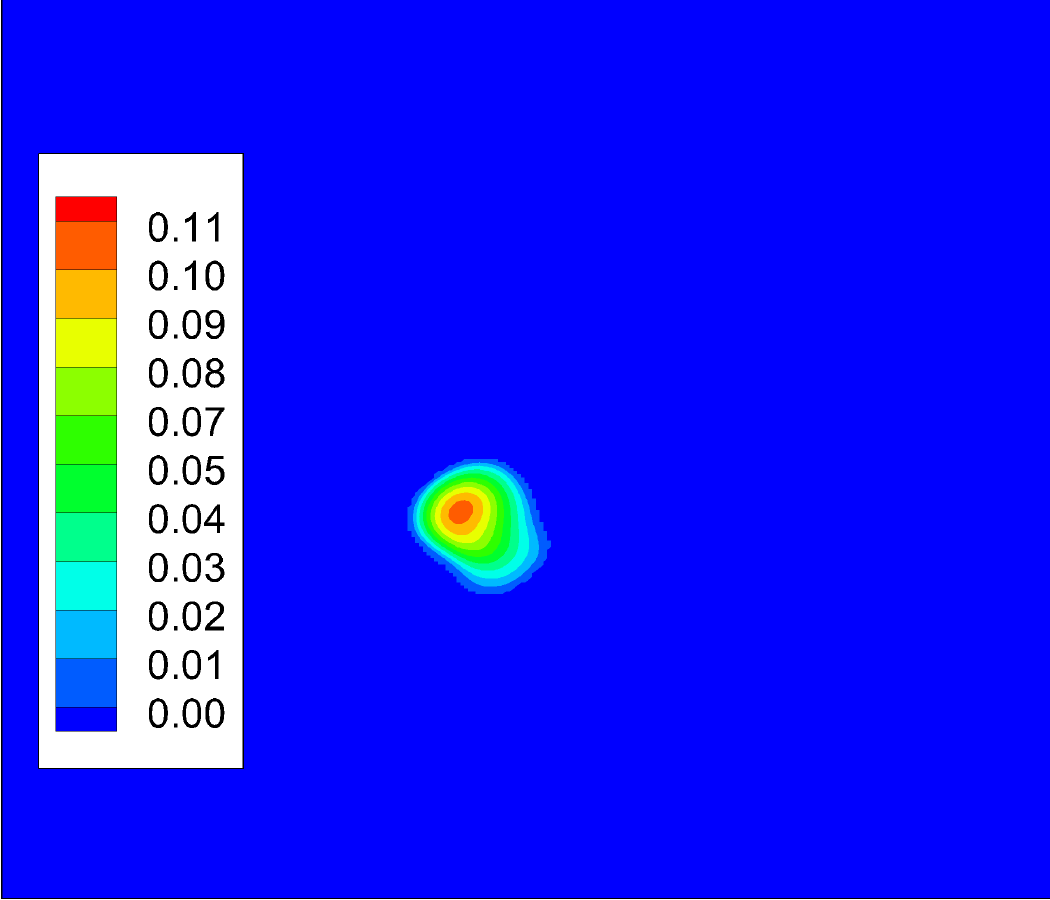}}\hfil
{\includegraphics[width=\tempwidth]{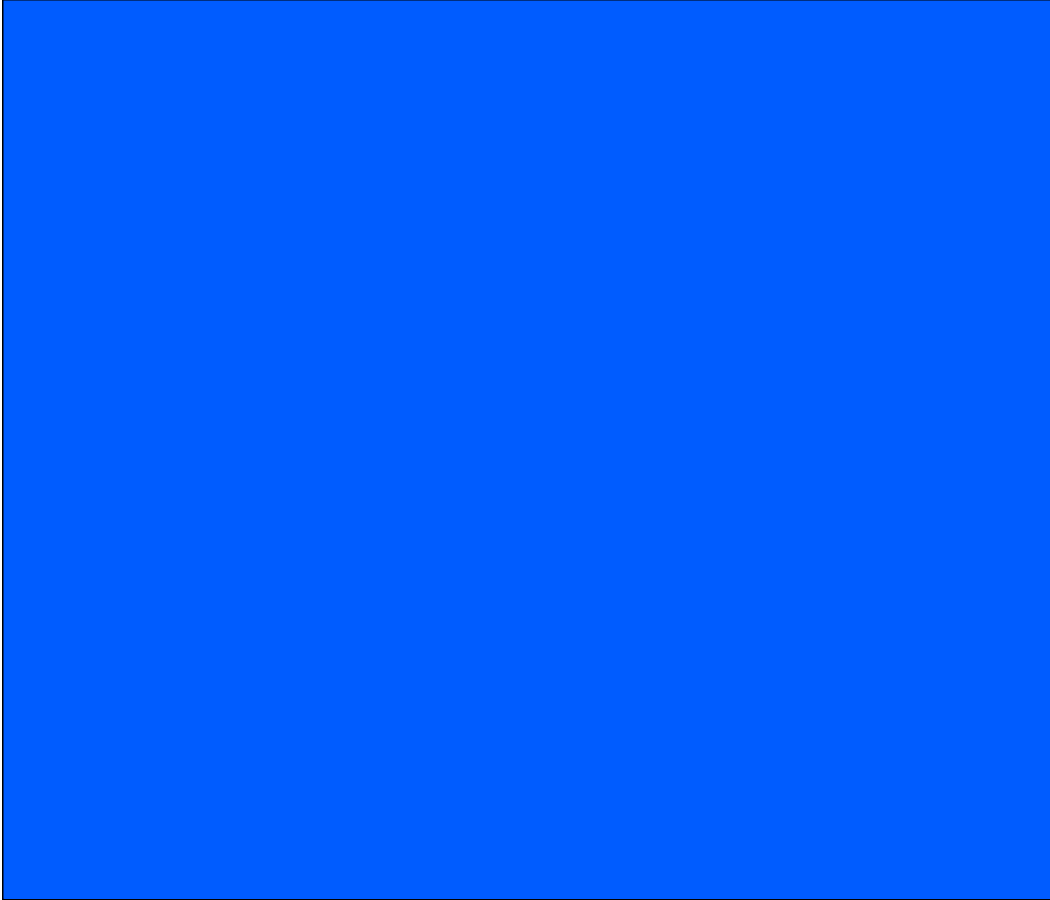}}\hfil
{\includegraphics[width=\tempwidth]{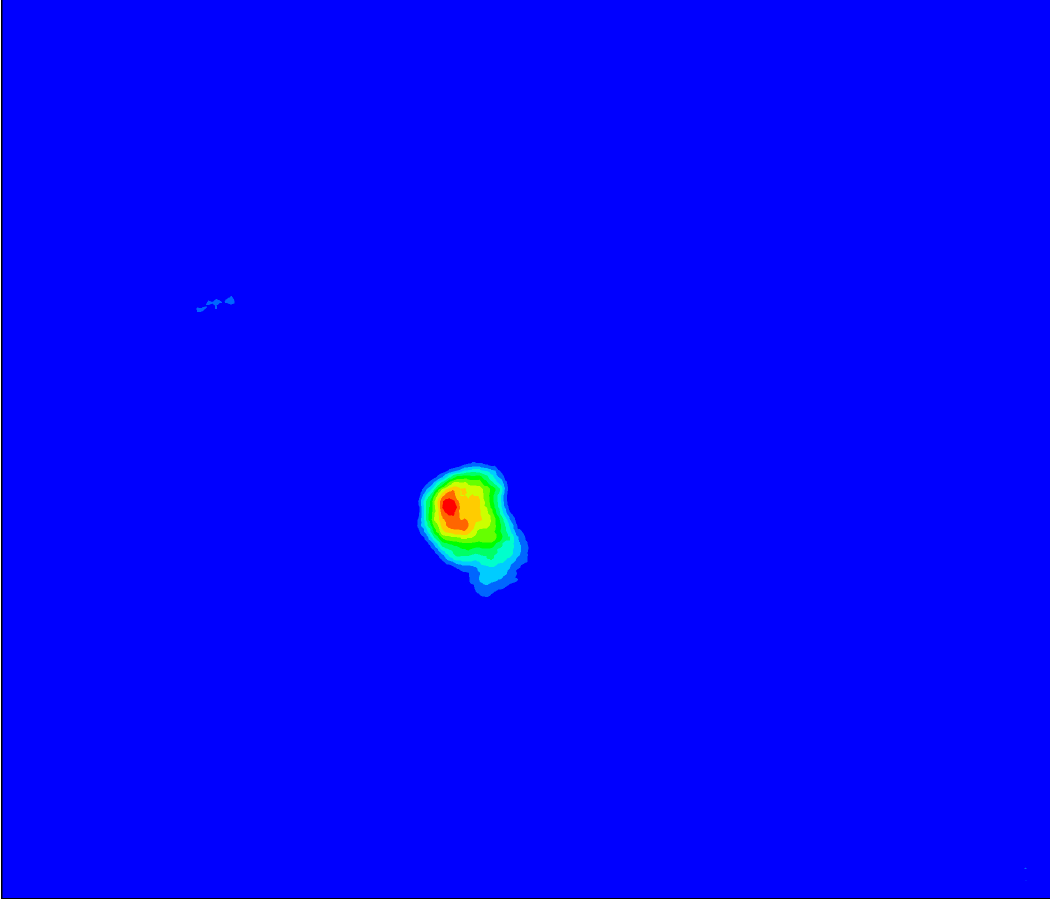}}\\
\rowname{Slice 4}
{\includegraphics[width=\tempwidth]{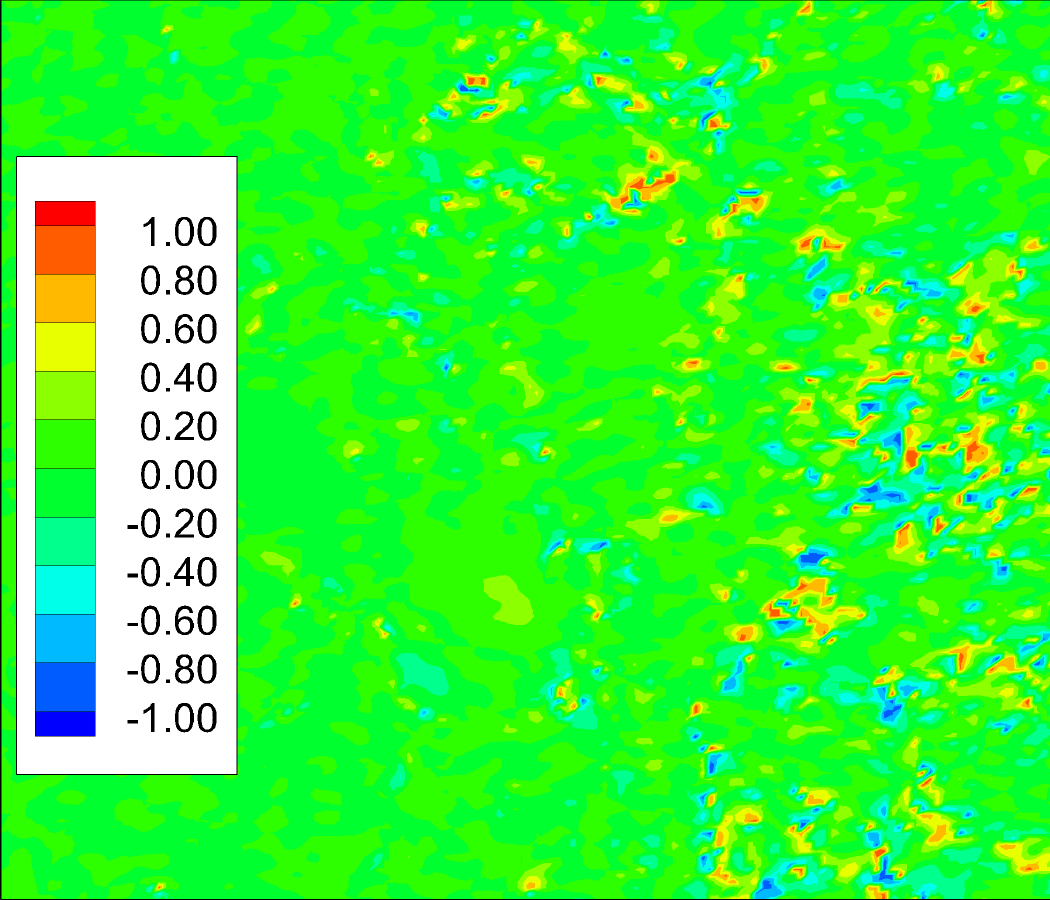}}\hfil
{\includegraphics[width=\tempwidth]{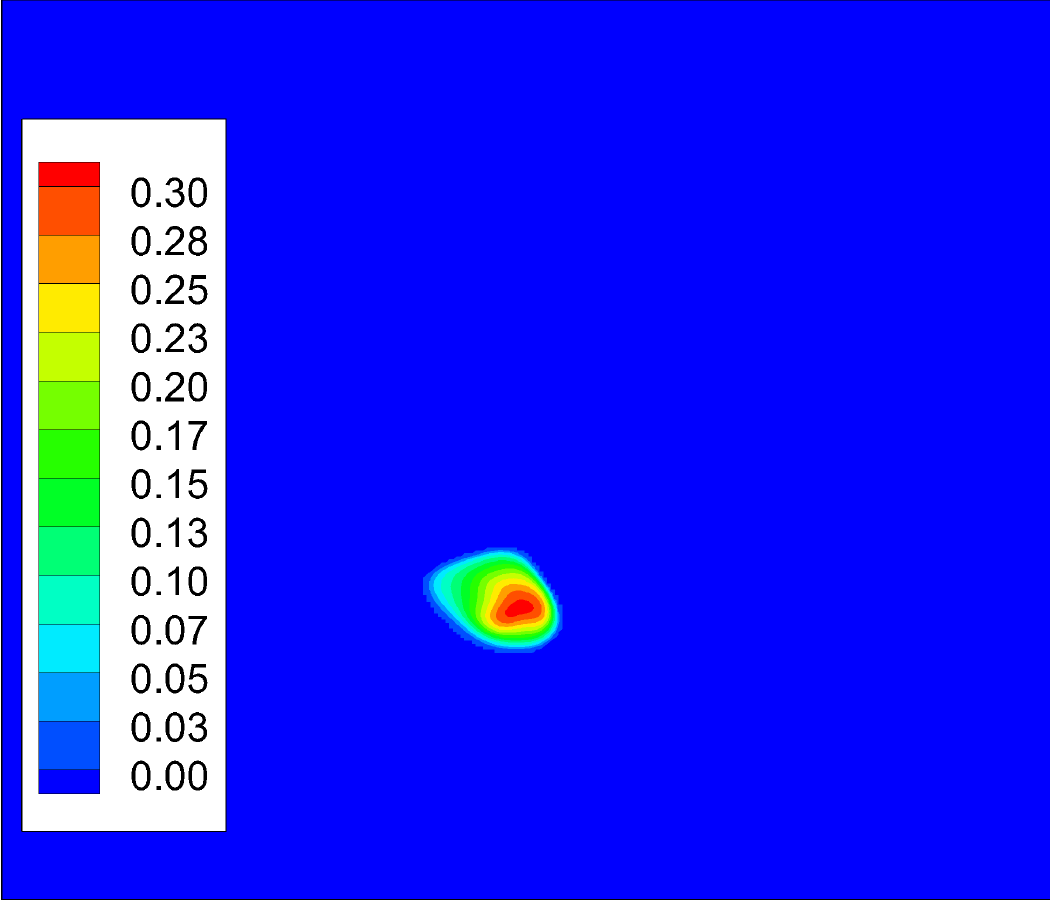}}\hfil
{\includegraphics[width=\tempwidth]{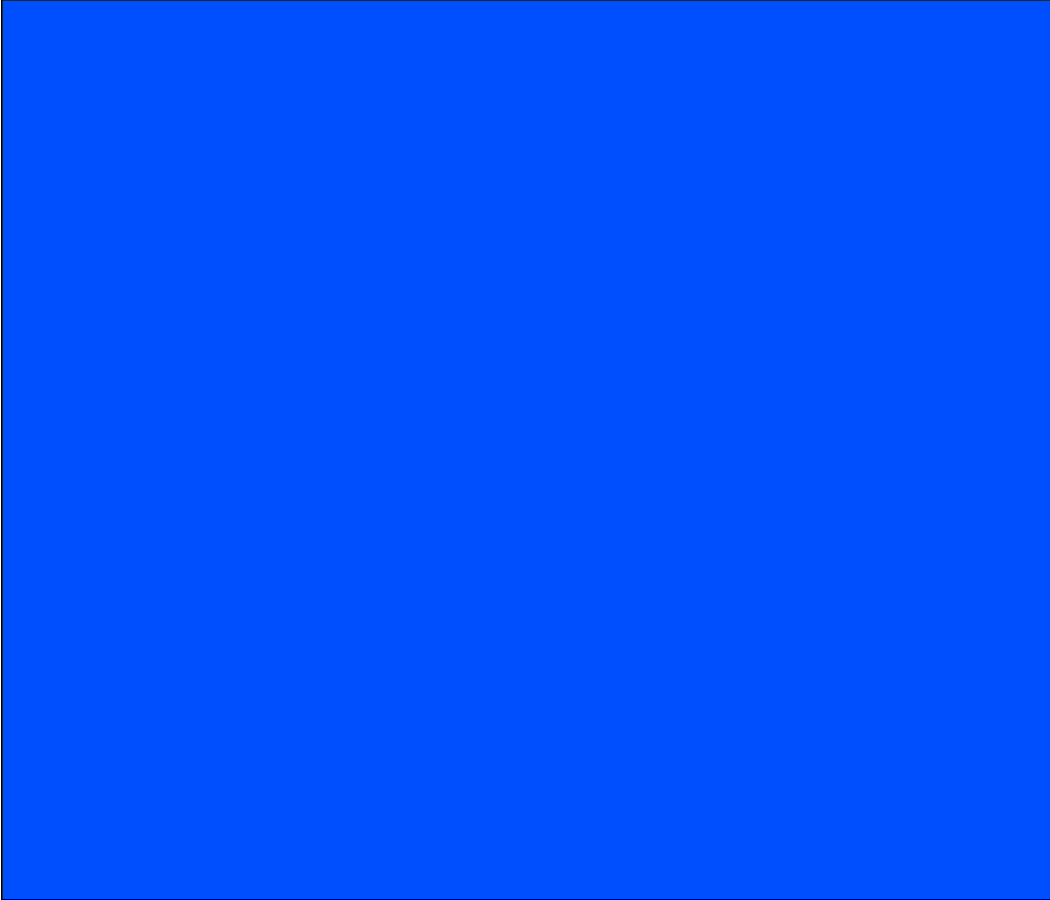}}\hfil
{\includegraphics[width=\tempwidth]{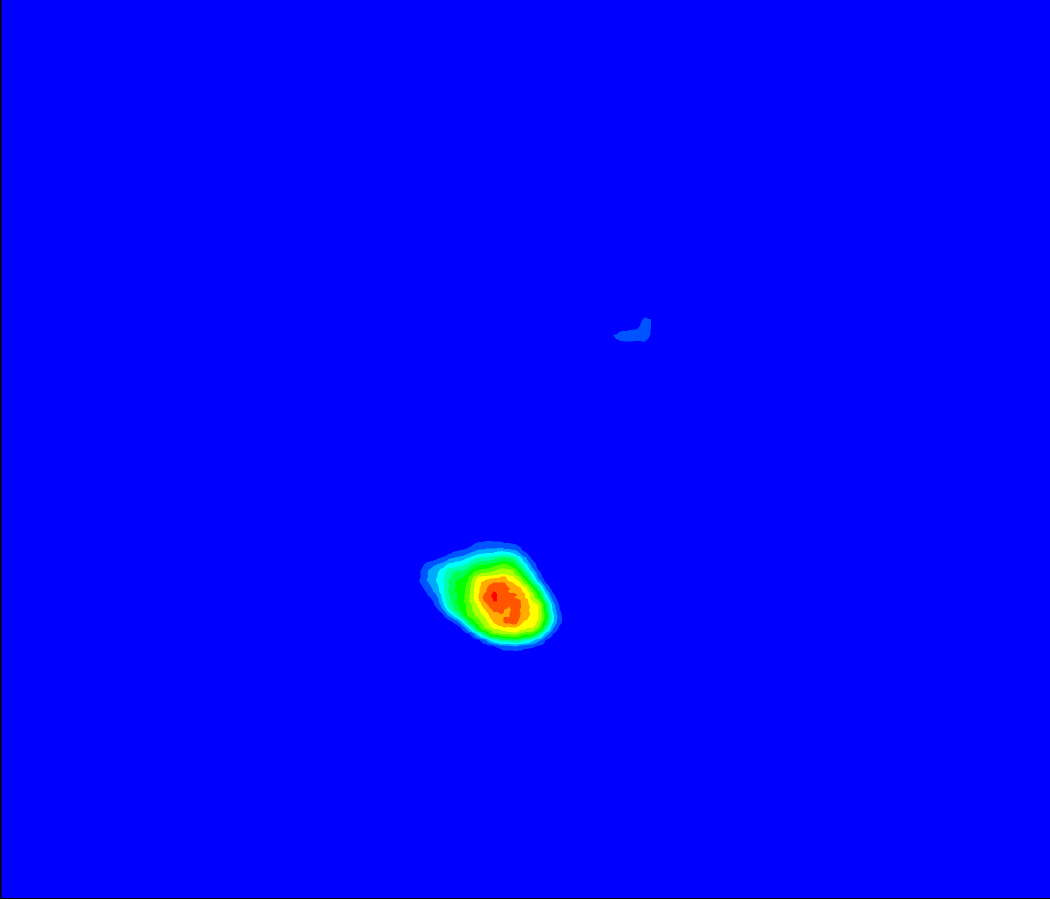}}\\
\caption{Comparison of the velocity contour predictions of the base model and the attention-based model. Slice 1: bifurcation region; Slice 2: ICA and ECA branches region; Slices 3 and 4: CCA region.}
\label{con1}
\end{figure}

In Figure~\ref{con1}, which displays velocity contours, the attention-based model performed better than the base model in identifying the velocity field, successfully capturing the complicated flow patterns that the base model missed. The attention model exhibited a significant ability to reconstruct velocity contours, although it showed discrepancies in both the values and locations of the predicted fields. These inconsistencies show that, even though the attention mechanism greatly boosts the model's performance in noisy settings, further adjustments may be needed to better match the actual data. The attention model's enhanced representation of velocity contours underscores its potential to advance super-resolution techniques in medical imaging applications.

The improvements obtained with the attention-based model are consistent with the gains reported for other deep learning-based super-resolution methods for 4D flow MRI. 4DFlowNet~\cite{ferdian20204dflownet} and its cerebrovascular extension~\cite{ferdian2023cerebrovascular}, the network of Rutkowski et al.~\cite{rutkowski2021enhancement} trained on CFD data, and SRflow~\cite{shit2022srflow} all reported clearly lower velocity errors than conventional interpolation when high-resolution targets derived from CFD were available. A direct numerical comparison with these studies is not possible, because they address different vascular territories (aorta, intracranial arteries), different resolution ratios, and different error definitions, and because the error in the present study is computed on the full volume rather than restricted to the vessel lumen. Our results extend the CFD-supervised paradigm to a cohort of 240 patient-specific carotid arteries and show that channel and spatial attention improve the robustness of the reconstruction to the measurement noise present in the low-resolution input.

Several limitations must be acknowledged. First, the CFD simulations served as ground truth, so the network learns to reproduce CFD velocity fields that themselves rest on modeling assumptions (rigid walls, Newtonian rheology, and inlet and outlet conditions derived from the MRI measurements); agreement with CFD is therefore not equivalent to agreement with the true in vivo flow. Second, only a resolution enhancement by a factor of two was investigated, and the performance at larger upsampling factors remains to be studied. Third, all data were acquired at a single center with a single scanner and protocol, so the generalizability to other scanners, field strengths, and vascular territories has not been established. Fourth, the evaluation was restricted to velocity errors; derived hemodynamic quantities such as wall shear stress, pressure gradients, and turbulent kinetic energy, which motivate the use of super-resolution, were not assessed. Fifth, the two models were compared with each other but not with conventional interpolation or with previously published super-resolution networks, and no independent in vivo validation, for example, against high-resolution 2D phase-contrast MRI or ultrasound measurements, was performed. Addressing these points, particularly the evaluation of derived hemodynamic quantities and validation against independent measurements, is the subject of future work.

\section{Conclusion}
This work addresses the issues of low resolution and noise in 4D flow MRI data, which restrict accurate blood flow analysis. To this end, we developed a deep learning architecture that integrates multi-scale feature extraction and attention mechanisms to improve spatial resolution and reduce noise. The main finding is that the attention-based model substantially outperformed the base model, achieving higher accuracy in reconstructing velocity fields and lower errors, thereby offering a more dependable method for assessing blood flow dynamics in cardiovascular diseases.

In the context of 4D flow MRI, the super-resolution model employing attention mechanisms performs well in improving image quality and detail; however, it has limitations. A major difficulty is the high noise level in the dataset, which may undermine the model's ability to produce reliable, accurate results. This noise can obscure fine details and produce artifacts that undermine overall image quality. A hybrid strategy that integrates the advantages of the super-resolution model with sophisticated denoising techniques may prove advantageous in addressing this issue. Moreover, it is essential to evaluate the temporal resolution of the imaging data, as enhancements in spatial resolution must not compromise temporal integrity. Balancing these aspects is crucial to enhancing the performance of super-resolution models in the clinical applications of 4D flow MRI.

\section*{Data availability statement}
The corresponding author will provide the datasets that support the conclusions of this article upon reasonable request.

\section*{Code availability statement}
The code used to build, train, and evaluate the models is available from the corresponding author upon reasonable request.

\section*{Conflict of interest}
The authors declare that they have no relevant financial or non-financial competing interests to disclose.

\section*{Ethical approval and consent to participate}
The original imaging studies from which the carotid 4D flow MRI dataset was derived were approved by the Ethics Committee of the University of Freiburg, Germany, and written informed consent was obtained from all participants. The work was performed in accordance with the Declaration of Helsinki. The present manuscript reports a secondary, retrospective analysis of fully anonymized data drawn from those original studies.

\section*{Consent for publication}
Not applicable, as no individual person's data are included in this manuscript.

\section*{Funding}
This research was financially supported by the Swiss National Science Foundation (SNSF) (grant \#205321L\_197189).

\section*{Acknowledgments}
The authors thank the Swiss National Supercomputing Center (CSCS) for providing computational resources under project ID s1290. Calculations for training the super-resolution models were performed on UBELIX (\url{https://www.id.unibe.ch/hpc}), the HPC cluster at the University of Bern.

\section*{Author contributions}
\textbf{Ali Mokhtari}: Conceptualization, Methodology, Software, Validation, Formal analysis, Investigation, Resources, Data curation, Writing (original draft), Writing (review and editing), Visualization.\\
\textbf{Dominik Obrist}: Funding acquisition, Supervision, Writing (review and editing).

\bibliographystyle{model1-num-names}

\bibliography{cas-refs}

\end{document}